\RequirePackage{ifpdf}
\ifpdf
\else
\PassOptionsToPackage{dvipdfmx}{graphicx}
\PassOptionsToPackage{dvipdfmx}{hyperref}
\fi

\pdfoutput=1

\documentclass[12pt,a4paper]{article}
\usepackage{jheppub}
\allowdisplaybreaks[4]
\usepackage{amsmath,bm}
\usepackage{subfigure}

\newcommand{\tr}{\textrm{Tr}}

\title{
Boundary driven turbulence on string worldsheet
}

\author[1]{Takaaki Ishii,}
\author[2]{Keiju Murata}
\author[3]{and Kentaroh Yoshida}
\affiliation[1]{Department of Physics, Rikkyo University, Nishi-Ikebukuro, Toshima, Tokyo 171-8501, Japan}
\affiliation[2]{Department of Physics, College of Humanities and Sciences, Nihon University, Sakurajosui, Tokyo 156-8550, Japan}
\affiliation[3]{Graduate School of Science and Engineering, Saitama University, 255 Shimo-Okubo, Sakura-ku, Saitama 338-8570, Japan}
\emailAdd{ishiitk@rikkyo.ac.jp}
\emailAdd{murata.keiju@nihon-u.ac.jp}
\emailAdd{kenyoshida@mail.saitama-u.ac.jp}

\abstract{%
We study the origin of turbulence on the string worldsheet with boundaries laid in anti de Sitter (AdS) spacetime. While the classical motion of a single closed string in AdS is integrable, it has recently been recognized that weak turbulence arises in the case of an open string suspended from the AdS boundary. In the open string case, it is necessary to impose boundary conditions on the worldsheet boundaries. We classify which boundary conditions preserve integrability. Based on this classification, we anticipate that turbulence may occur on the string worldsheet if integrability is not guaranteed by the boundary conditions. Numerical investigations of the classical open-string dynamics support that turbulence occurs when the boundary conditions are not integrable.
}

\preprint{RUP-23-18, STUPP-23-267}

\begin{document}
\maketitle

\section{Introduction}

Some turbulent behaviors of a single fundamental string in an asymptotically AdS spacetime have been studied from various aspects~\cite{Mikhailov:2003er,Callebaut:2015fsa,Vegh:2015yua,Ishii:2015wua,Ishii:2015qmj,Ali-Akbari:2015ooa,Vegh:2016hwq,Gubser:2016wno,Ishii:2016rlk,Vegh:2018dda,Hashimoto:2018fkb,Akutagawa:2019awh}. In particular, the work~\cite{Ishii:2015wua} studied the classical dynamics of an open string whose endpoints are attached to the boundary of Poincar\'{e} AdS and revealed the weak turbulence in the string dynamics. That is, an energy cascade from large to small scales occurs on the string worldsheet and eventually the energy spectrum obeys the power law. (See also the weak turbulence in an Einstein-scalar system~\cite{Bizon:2011gg} and the D3/D7 system~\cite{Hashimoto:2014xta,Hashimoto:2014dda}.) The turbulence is a typical phenomenon in non-integrable systems. However, the classical dynamics of a closed string in AdS is integrable due to the existence of an infinite number of conserved charges \cite{Mandal:2002fs}. What is the origin of the turbulence of the AdS string? Our anticipation is that this turbulence should be intrinsic to the open string case, and the breakdown of integrability due to the choice of boundary conditions may cause turbulence.

In \cite{Ishii:2015wua}, the classical dynamics of the open string was studied for a probe string dual to the Wilson loop giving rise to quark-antiquark potential \cite{Rey:1998ik,Maldacena:1998im}. The string is hanging in the Poincar\'{e} AdS$_5$ from the boundary. Perturbation is induced on the string by moving the endpoints momentarily and nonlinear time evolution was calculated numerically. For the simplest kind of perturbation, the string dynamics is confined in the AdS$_3$ slice of AdS$_5$ and so the AdS$_3$ consideration is sufficient to find the weak turbulence. Hence, in this paper, we will focus on the string dynamics in AdS$_3$. In addition, we will adopt the global coordinates to borrow techniques of integrable sigma models.

The string dynamics in AdS$_3$ can be described by the principal chiral model (PCM) whose target space is an $SL(2,R)$ group manifold. (For a review, for example, see \cite{Yoshidabook}.) It has been shown that the PCM {\it without} boundaries is integrable~\cite{Evans:1999mj}. Meanwhile, an open string has two endpoints and its dynamics is described by the PCM {\it with} boundaries. In the presence of boundaries, the integrability may be broken depending on the choice of boundary conditions, but under certain conditions, it is still possible to construct an infinite number of conserved charges and maintain integrability as shown in a series of  works~\cite{MacKay:2001bh,Delius:2001he,Mann:2006rh,Dekel:2011ja,MacKay:2011zs}. 

In this paper, we will classify integrable boundary conditions for the open string in AdS$_3$. However, this story has a downside that we can give only the sufficient condition to preserve the integrability. Hence, it is not possible to conclude non-integrability when the integrable boundary conditions are not satisfied. To test non-integrablity depending on boundary conditions, we will solve the dynamics of the open string numerically and show the existence of the weak turbulence when the boundary conditions do not guarantee the integrability. 

For numerical calculations, we consider an open string connecting two points on the boundary of the global AdS$_3$. In particular, when the string connects the antipodal points of the boundary $S^1$, the open string is ``straight'' in the AdS bulk with the AdS$_2$ induced metric on the string worldsheet. For the straight string, nonlinear waves can be produced \cite{Mikhailov:2003er}. For these waves, the reflection by the AdS boundary was not fully analyzed. In this paper, we investigate nonlinear dynamics of the open string by taking account of the presence of the boundaries on the string worldsheet.

This paper is organized as follows. Section~\ref{FstringAdS3} introduces the equations of motion of a fundamental string in AdS$_3$ with double null worldsheet coordinates. The conserved currents associated with the isometry group of AdS$_3$ are also obtained. Section~\ref{PCMsection} is a brief review of the PCM with the target space $SL(2,R)$. It is shown that an infinite number of non-local conserved charges are generated by following the standard method. We further introduce boundaries in the PCM and show that we can obtain an infinite number of conserved charges depending on the choices of boundary conditions.
In section~\ref{integrableAdSstring}, we identify integrable boundary conditions for the open string in AdS$_3$ from the integrability conditions of the PCM with boundaries. In section~\ref{sec:nonlinear}, we introduce the formulation for the numerical calculations of the nonlinear dynamics of classical open string in the global AdS$_3$, where the string endpoints are fixed on the boundary. In section~\ref{sec:results}, results of the numerical calculations are shown. We find the weak turbulence, that gives strong evidence of the non-integrability. The last section is devoted to conclusion and discussion. Appendix~\ref{examineall} provides detailed analysis of the integrability condition of the open string that is omitted in the main text.

\section{Fundamental string in AdS$_3$}
\label{FstringAdS3}

In this section, we shall introduce the Nambu-Goto action describing a sting moving in AdS$_3$.

\subsection{AdS$_3$ spacetime}
The metric of the space $R^{2,2}$ is written as
\begin{equation}
 ds^2=\eta_{MN}dX^M dX^N \ ,
\label{dsR22}
\end{equation}
where $M,N=0,1,2,3$ and $\eta_{MN}=\textrm{diag}(-1,1,1,-1)$.
The metric of AdS$_3$ is realized as the induced metric on the equidistance surface in $R^{2,2}$: 
\begin{equation}
 \eta_{MN}X^M X^N=-\ell_{\textrm{AdS}}^2=-1\ ,
\label{Xsq}
\end{equation}
where $\ell_{\textrm{AdS}}$ is the AdS radius, and we set $\ell_{\textrm{AdS}}=1$ throughout this paper.
The above equation can be solved by
\begin{equation}
 X^0+iX^3=\frac{1+r^2}{1-r^2}\, e^{it}\ ,\quad X^1+iX^2=\frac{2r}{1-r^2}\,e^{i\theta}\ .
\label{Xglobal}
\end{equation}
The coordinates $(t,r,\theta)$ cover the global AdS$_3$, where the metric is given by\footnote{With a new radial coordinate
$\rho=2r/(1-r^2)$, the AdS$_3$ metric becomes a familiar form: $ds^2=-(\rho^2+1)dt^2+d\rho^2/(\rho^2+1)+\rho^2d\theta^2$.
}
\begin{equation}
ds^2=-\left(\frac{1+r^2}{1-r^2}\right)^2dt^2 +\frac{4}{(1-r^2)^2}(dr^2+r^2d\theta^2)\ .
\label{dsglobal}
\end{equation}
Here, the surface $r=1$ corresponds to the AdS boundary.
The center of AdS$_3$ ($r=0$) is the coordinate singularity in this coordinate system, which is inconvenient for numerical calculations. To avoid this coordinate singularity, it is convenient to introduce the ``Cartesian'' coordinates defined as
\begin{equation}
x \equiv r\cos\theta \ , \qquad y \equiv r\sin\theta \ .
\label{xy_coordinates}
\end{equation}
Then, the metric \eqref{dsglobal} is written as
\begin{equation}
 ds^2=-\left(\frac{1+x^2+y^2}{1-x^2-y^2}\right)^2dt^2 +\frac{4}{(1-x^2-y^2)^2}(dx^2+dy^2)\ .
\label{AdScartes}
\end{equation}
The coordinates are smooth at the origin $x=y=0$.

It is also convenient hereafter to take the coordinates $(\bar{t},\bar{r},\bar{\theta})$ 
defined by the following relations:\footnote{For example, see \cite{Duff:1998cr}.} 
\begin{equation}
 \begin{split}
  X^0+iX^3&=\frac{1}{\sqrt{2}}\, e^{i\bar{t}/2}\left\{\cosh \left(\frac{\bar{\theta}+\bar{r}}{2}\right)+ i\cosh \left(\frac{\bar{\theta}-\bar{r}}{2}\right) \right\}\ , \\
X^1+iX^2&=\frac{1}{\sqrt{2}}\, e^{i\bar{t}/2}\left\{\sinh \left(\frac{\bar{\theta}+\bar{r}}{2}\right)+ i\sinh \left(\frac{\bar{\theta}-\bar{r}}{2}\right) \right\}\ .
 \end{split}
\label{Xtwist}
\end{equation}
It is easy to check that this parametrization also satisfies the condition (\ref{Xsq}). Then, the metric of AdS$_3$ is written as 
\begin{equation}
 ds^2=\frac{1}{4}\{
-\cosh^2\bar{r}\,d\bar{t}^2+d\bar{r}^2+(d\bar{\theta}+\sinh \bar{r}\, d\bar{t})^2
\}\ .
\label{AdS3twist}
\end{equation}
The coordinate transformation between ($t,r,\theta$) and ($\bar{t},\bar{r},\bar{\theta}$) is complicated, but it can be obtained explicitly by using Eqs.(\ref{Xglobal}) and (\ref{Xtwist}).

\subsection{Equations of motion}

Let us consider the classical dynamics of a single string moving in AdS$_3$. 
In the following, we will derive the string equations of motion without specifying explicit AdS$_3$ coordinates.

The Nambu-Goto action is given by
\begin{equation}
S=-\frac{1}{2\pi\alpha'}\int d^2\sigma\, \sqrt{-h}\ ,
\label{NambuGotoAction}
\end{equation}
where $h_{ab}$ is the induced metric on the string worldsheet.
The string coordinates in $R^{2,2}$ are written as functions of the worldsheet coordinates $(\sigma^+,\sigma^-)$,
\begin{equation}
 X^M=X^M(\sigma^+,\sigma^-)\ .
\end{equation}
The components of the induced metric are
\begin{equation}
 h_{\pm\pm}=\eta_{MN}\partial_{\pm}X^M \partial_\pm X^N\ ,\quad
 h_{+-}=\eta_{MN}\partial_+ X^M \partial_- X^N\ ,
\end{equation}
where $\partial_{\pm}\equiv\partial/\partial \sigma^\pm$.

The diffeomorphism invariance enables us to impose the double null condition: 
\begin{equation}
 h_{\pm\pm}=0\ .
\label{virasoro}
\end{equation}
Then, we find $\sqrt{-h}=\sqrt{h_{+-}^2-h_{++}h_{--}}=-h_{+-}$. Note that $h_{+-}$ is negative when $\partial_+$ and $\partial_-$ are future directed. 
Thus, the Nambu-Goto action can be simply written as 
\begin{align}
 S&=\frac{1}{2\pi\alpha'}\int d\sigma^+d\sigma^- \,
  \partial_+ X^M \partial_- X^N \eta_{MN} \ ,
  \label{Sstring0}
\end{align}
where the equidistance condition~(\ref{Xsq}) is implicitly imposed.
This theory is known as the nonlinear sigma model of $SO(2,2)$. We still have residual degrees of freedom of the conformal transformation, 
\begin{equation}
 \sigma^\pm =\sigma^\pm( \sigma'{}^\pm)\ .
\label{resid}
\end{equation}

Using the worldsheet metric $\gamma^{ab}$ $(a,b=+,-)$, (\ref{Sstring0}) can be written as 
\begin{align}
 S &=\frac{1}{4\pi\alpha'}\int d\sigma^+d\sigma^- \,
  \gamma^{ab}\partial_a X^M \partial_b X^N \eta_{MN} \ , \qquad
  \gamma^{ab}=\begin{pmatrix}
      0 & 1 \\ 1 & 0
  \end{pmatrix}
\ .
\end{align}
Then, the worldsheet energy momentum tensor is given by
\begin{equation}
T_{ab} = \partial_a X^M \partial_b X^N \eta_{MN} - \frac{1}{2} \gamma_{ab} \partial_c X^M \partial^c X^N \eta_{MN} \ .
\end{equation}
This satisfies
\begin{equation}
    T_{++}=h_{++}=0 \ ,  \quad T_{--}=h_{--}=0 \ ,  \quad T_{+-}=0 \ ,
\end{equation}
where the first two relations follow from the double null condition \eqref{virasoro}, and the last equation holds identically.

As in Eqs.(\ref{Xglobal}) and (\ref{Xtwist}), we can express $X^M$ by unconstrained coordinates $x^{\mu}$~($\mu=0,1,2$) as $X^M = X^M(x^\mu)$. Then, the above action can be expressed as
\begin{equation}
 S=\frac{1}{2\pi\alpha'}\int d\sigma^+d\sigma^- \,
  g_{\mu\nu}(x^\mu(\sigma^+,\sigma^-))\partial_+ x^\mu \partial_- x^\nu\ ,
\label{Sstring1}
\end{equation}
where $g_{\mu\nu}=\eta_{MN}\partial_\mu X^M \partial_\nu X^N$ is the metric of AdS$_3$.

In the action~(\ref{Sstring0}), we can explicitly realize the equidistance condition~(\ref{Xsq}) by using the Lagrange multiplier $\lambda$ as follows: 
\begin{equation}
 S=\frac{1}{2\pi\alpha'}\int d\sigma^+d\sigma^- \left[
  \eta_{MN}\partial_+ X^M \partial_- X^N+\lambda(\eta_{MN}X^M X^N+1)\right]\ .
\end{equation}
Then the equations of motion are given by 
\begin{equation}
\partial_+ \partial_- X^M=\lambda\, X^M \quad \mbox{with} \quad  \eta_{MN}X^M X^N=-1 \ .
\end{equation}
By multiplying $X_M$ to the first equation and summing over $M$, we obtain
$\lambda=-X_M \partial_+ \partial_- X^M$. Then, eliminating $\lambda$ leads to 
\begin{equation}
 \partial_+ \partial_- X^M=- X^M X_N \partial_+ \partial_- X^N 
 \quad \mbox{with} \quad  \eta_{MN}X^M X^N=-1\ .
\label{EOM}
\end{equation}

In the following analysis, it is convenient to introduce new coordinates $(\tau,\sigma)$ by
\begin{equation}
 \tau=\sigma^+ + \sigma^-\ ,\quad \sigma=\sigma^+-\sigma^-\ .
\label{tausigma}
\end{equation}
Then, Eq.(\ref{EOM}) is written as
\begin{equation}
 \partial^2 X^M=- X^M X_N \partial^2 X^N\ \quad \mbox{with} \quad  \eta_{MN}X^M X^N=-1\ , 
\label{EOM2}
\end{equation}
where $\partial^2=-\partial^2_\tau+\partial_\sigma^2$.

\subsection{Isometry group of AdS$_3$ and  conserved currents}

The AdS$_3$ spacetime has the $SO(2,2)$-isometry. The Killing vectors are given
by 
\begin{equation}
 \xi_{MN}=X_M \frac{\partial}{\partial X^N}- X_N \frac{\partial}{\partial X^M}\ ,
\label{xiKilling}
\end{equation}
where $X_M=\eta_{MN}X^N$, and each of the $\xi_{MN}$ generates a ``rotation'' in the $(X^M,X^N)$-plane. 
The conserved currents on the worldsheet, which are associated with the Killing vectors~(\ref{xiKilling}), are given by
\begin{equation}
 (A_{MN})_a = 2(X_M \partial_a X_N-X_M \partial_a X_N) \ ,
\label{J1}
\end{equation}
where $a,b=\tau,\sigma$. Using the equations of motion~(\ref{EOM2}), one can directly check that the above quantity is conserved, 
\begin{equation}
 \partial^a (A_{MN})_a=0\ .
\label{dJ0}
\end{equation}
We will use the 1-form $A_{MN}=(A_{MN})_a\, d\sigma^a$ in the following analysis.

Because $SO(2,2)\simeq SL(2)_L \times SL(2)_R$, we can obtain the generators of $SL(2)_{L,R}$ by taking appropriate linear combinations of the Killing vectors~(\ref{xiKilling}).
The generators of $SL(2)_{L}$ and $SL(2)_R$ are denoted by $L_i$ and $\bar{L}_i$ ($i=0,1,2,3$), respectively. 
They are explicitly written as 
\begin{equation}
\begin{split}
&L_0 = (\xi_{03}-\xi_{12})/2\ ,\quad
L_1 = (-\xi_{13}+\xi_{02})/2\ ,\quad
L_2 = (\xi_{01}+\xi_{23})/2\ ,\\
&\bar{L}_0 = (-\xi_{03}-\xi_{12})/2\ ,\quad \bar{L}_1 = (-\xi_{01}+\xi_{23})/2\ ,\quad \bar{L}_2 = (\xi_{13}+\xi_{02})/2\ .
\end{split}
\end{equation}
The commutation relations are given by
\begin{equation}
 \begin{split}
 &[L_0,L_1]=-L_2\ ,\quad [L_0,L_2]=L_1\ ,\quad [L_1,L_2]=L_0\ ,\\
&[\bar{L}_0,\bar{L}_1]=-\bar{L}_2\ ,\quad [\bar{L}_0,\bar{L}_2]=\bar{L}_1\ ,\quad [\bar{L}_1,\bar{L}_2]=\bar{L}_0\ .
\end{split}
\end{equation}
The conserved currents associated with $L_i$ and $\bar{L}_i$ are 
\begin{equation}
\begin{split}
&I^0 = (A_{03}-A_{12})/2\ ,\quad
I^1 = (-A_{13}+A_{02})/2\ ,\quad
I^2 = (A_{01}+A_{23})/2\ ,\\
&J^0 = (-A_{03}-A_{12})/2\ ,\quad
J^1 = (-A_{01}+A_{23})/2\ ,\quad
J^2 = (A_{13}+A_{02})/2\ .
\end{split}
\label{IJandA}
\end{equation}
We will see that, for closed strings or open strings under certain boundary conditions, an infinite number of conserved charges can be generated from $J^A$ and $I^A$ ($A=0,1,2$).

\section{Principal chiral model}
\label{PCMsection}

In this section, we use the PCM with $SL(2,R)$ target space and discuss the conditions of integrability in the presence of boundaries on the worldsheet.

\subsection{Symmetry and conserved currents}

It is known that the dynamics of the string in AdS$_3$ is described by the PCM whose target space is given by $SL(2,R)$. Let us here provide a short review of the PCM. The action of the PCM is given by  
\begin{equation}
 S=-\frac{1}{2}\int d^2\bm{\sigma}\, \eta^{ab} \tr(J_a J_b)=-\frac{1}{2}\int d^2\bm{\sigma}\, \eta^{ab} \tr(I_a I_b)\  .
\label{PCM}
\end{equation}
where $\bm{\sigma}=(\tau,\sigma)$ and
\begin{equation}
 J_a \equiv g^{-1}(\bm{\sigma})\partial_a g(\bm{\sigma})\ ,\quad 
 I_a \equiv -\partial_a g(\bm{\sigma}) g^{-1}(\bm{\sigma}) ,\quad g(\bm{\sigma})\in SL(2,R)\  .
\label{Jdef}
\end{equation}
In the PCM, the dynamical variable is a group element $g(\bm{\sigma}) \in SL(2,R)$. 
The equation of motion of the PCM is simply given by
\begin{equation}
 \partial^a J_a = \partial^a I_a = 0\ .
\label{JIconsv}
\end{equation}
This shows that $J_a$ and $I_a$ are conserved currents. 
We also write these currents by 1-forms as 
\begin{align}
 J&=J_a\, d\sigma^a=g^{-1}dg\ , \\
 I&=I_a\, d\sigma^a=-dg \cdot g^{-1}\ , 
\end{align}
which are referred to as the left and right invariant 1-forms, respectively.

Note that $J$ and $I$ take values into the associated Lie algebra $sl(2,R)$ by definition and can be expanded as 
\begin{equation}
 J=J_A T^A=J_{aA}T^A d\sigma^a\ ,\quad I=I_A T^A=I_{aA}T^A d\sigma^a\ .
\end{equation}
where $T^A$ ($A=0,1,2$) are the generators of $SL(2)$ satisfying 
\begin{equation}
 [T^0,T^1]=-T^2\ ,\quad [T^1,T^2]=T^0\ ,\quad [T^2,T^0]=-T^1\ .  
\end{equation}
The fundamental representation of $T^A$'s is given by 
\begin{equation}
 T^0 = \frac{i}{2}\sigma_1\ ,\quad T^1 = \frac{1}{2}\sigma_2\ ,\quad T^2 = \frac{1}{2}\sigma_3\ ,
\end{equation}
by using the Pauli matrices $\sigma_{1,2,3}$. Then the orthogonality condition is given by 
\begin{equation}
 \tr(T^AT^B) = \frac{1}{2}\textrm{diag}(-1,1,1) \equiv \gamma^{AB}\ .
\end{equation}
The indices $A,B,\cdots$ are raised and lowered by using $\gamma^{AB}$ and $\gamma_{AB}=2\,\textrm{diag}(-1,1,1)$.
For example, $J^A\equiv \gamma^{AB}J_B$ and $I^A\equiv \gamma^{AB}I_B$. 

The structure constants are defined by $[T^A,T^B]=f^{AB}{}_C T^C$. For $SL(2)$, nonzero components contain $f^{01}{}_2=-1$, $f^{12}{}_0=1$ and $f^{20}{}_1=-1$. These are simply expressed as
\begin{equation}
 f_{ABC}\equiv \gamma_{AD}\gamma_{BE}f^{DE}{}_C = 4\epsilon_{ABC}\ ,
\end{equation}
where $\epsilon_{ABC}$ is the totally antisymmetric tensor with $\epsilon_{012}=1$.

One can check that $J$ and $I$ are invariant under global transformations $g\to g_L g$ and $g\to g g_R$, respectively ($g_L, g_R\in SL(2)$). This is the origin of the names the left and right invariant 1-forms. 
It follows that the action~(\ref{PCM}) is invariant under
\begin{equation}
 g(\bm{\sigma})\to g_L \, g(\bm{\sigma})\,  g_R\ .
\end{equation}
Namely, the PCM~(\ref{PCM}) has the global symmetry $SL(2)_L\times SL(2)_R\simeq SO(2,2)$, that is the same as the isometry group of AdS$_3$. Then $J$ and $I$ are conserved current with respect to $SL(2,R)_R$ and $SL(2,R)_L$, respectively.

We can also prove that the ``field strength'' with respect to $J$ and $I$ is zero:
\begin{equation}
 \partial_a J_b-\partial_b J_a + [J_a,J_b]=0\ ,\quad
 \partial_a I_b-\partial_b I_a + [I_a,I_b]=0\ .
\label{flatcond}
\end{equation}
The above equations are called the {\textit{flatness conditions}} and are essential to generate an infinite number of conserved charges.
Using the conservation equations of $J$ and $I$ (\ref{JIconsv}), the flatness conditions are written as
\begin{equation}
\partial_\pm J_{\mp}=-\frac{1}{2}[J_\pm,J_\mp] ,\quad 
\partial_\pm I_{\mp}=-\frac{1}{2}[I_\pm,I_\mp]\ ,
\label{flatcond2}
\end{equation}
where $\sigma^\pm$ are the double null coordinates defined in Eq.(\ref{tausigma}).

\subsection{AdS$_3$ metric from principal chiral model}

We can derive the classical string action~(\ref{Sstring1}) from the PCM~(\ref{PCM}). Let us parameterize $g(\bm{\sigma})\in SL(2,R)$ as
\begin{equation}
 g(\bm{\sigma})=\exp[\bar{t}(\bm{\sigma})T^0]\exp[\bar{r}(\bm{\sigma})T^2]\exp[\bar{\theta}(\bm{\sigma})T^1]\ .
\end{equation}
Then, by direct calculations, the invariant 1-forms are given by
\begin{equation}
\begin{split}
J&=\left[\frac{1}{2}\left\{\cosh(\bar{\theta}+\bar{r})+\cosh(\bar{\theta}-\bar{r})\right\}d\bar{t}-\sinh\bar{\theta} \, d\bar{r}\right]T^0
+(\sinh \bar{r} \, d\bar{t}+d\bar{\theta})T^1\\
&\hspace{3.5cm}+\left[-\frac{1}{2}\left\{\sinh(\bar{\theta}+\bar{r})+\sinh(\bar{\theta}-\bar{r})\right\}d\bar{t}+\cosh\bar{\theta} \, d\bar{r}\right]T^2\ ,\\
I&=(-d\bar{t}+\sinh\bar{r}\, d\bar{\theta})T^0
-(\sin \bar{t} \, d\bar{r}+\cosh\bar{r}\cos\bar{t}\, d\bar{\theta})T^1\\
&\hspace{7cm}+(-\cos \bar{t} \, d\bar{r}+\cosh\bar{r}\sin\bar{t}\, d\bar{\theta})T^2\ .
\end{split}
\end{equation}
From the above expressions, we obtain
\begin{equation}
\begin{split}
 \frac{1}{2}\tr(JJ)=\frac{1}{2}\tr(II)
=\frac{1}{4}\left\{
-\cosh^2\bar{r}\,d\bar{t}^2+d\bar{r}^2+(d\bar{\theta}+\sinh \bar{r}\, d\bar{t})^2 \right\}\ .
\end{split}
\label{TrJJ}
\end{equation}
This is the  metric of AdS$_3$ as given in Eq.(\ref{AdS3twist}). 
From Eq.(\ref{TrJJ}), the PCM action~(\ref{PCM}) is written as
\begin{equation}
 S=-\int d^2\bm{\sigma}\, \eta^{ab} g_{\mu\nu}(x(\bm{\sigma})) \frac{\partial x^\mu}{\partial \sigma^a} \frac{\partial x^\nu}{\partial \sigma^b}\  .
\end{equation}
This is nothing but the string action~(\ref{Sstring1}) written in the  $(\tau,\sigma)$ coordinates. 

Note that in the string case, the constraints~(\ref{virasoro}) must be satisfied as additional conditions at the initial surface and boundaries of the worldsheet. These are not discussed in the usual context of the PCM 
because two-dimensional gravity is not coupled in the usual PCM setup where two-dimensional flat space is prepared from the beginning.

\subsection{Integrability for closed strings}

We show the classical integrability of a single closed string propagating in AdS$_3$~\cite{Mandal:2002fs}. (The arguments in this subsection are based on Ref.\cite{Mann:2006rh}. See also Ref.\cite{Yoshidabook} for the review of the construction of the Lax pair and monodromy matrix of the PCM.)

Let us introduce Lax pairs $\mathcal{L}_\pm$ and $\bar{\mathcal{L}}_\pm$ as
\begin{equation}
 \mathcal{L}_\pm(\bm{\sigma};\lambda) = \frac{J_\pm(\bm{\sigma})}{1\pm \lambda}\ ,\quad \bar{\mathcal{L}}_\pm(\bm{\sigma};\lambda) = \frac{I_\pm(\bm{\sigma})}{1\pm \lambda}\ ,
\label{Laxdef}
\end{equation}
where $\lambda$ is a constant called the spectral parameter. 
The above equations are also written as  
\begin{equation}
 \mathcal{L}_\tau=\frac{J_\tau-\lambda J_\sigma}{1-\lambda^2}\ ,\quad \mathcal{L}_\sigma=\frac{J_\sigma-\lambda J_\tau}{1-\lambda^2}\ ,\quad 
\bar{\mathcal{L}}_\tau=\frac{I_\tau-\lambda I_\sigma}{1-\lambda^2}\ ,\quad \bar{\mathcal{L}}_\sigma=\frac{I_\sigma-\lambda I_\tau}{1-\lambda^2} \ ,
\end{equation}
where $\mathcal{L}_\pm=\mathcal{L}_\tau \pm \mathcal{L}_\sigma$ and $\bar{\mathcal{L}}_\pm=\bar{\mathcal{L}}_\tau \pm \bar{\mathcal{L}}_\sigma$.
Under Eq.(\ref{flatcond2}), the flatness conditions for $\mathcal{L}_a$ and $\bar{\mathcal{L}}_a$ are satisfied as 
\begin{equation}
\partial_a \mathcal{L}_b - \partial_b \mathcal{L}_a +[\mathcal{L}_a,\mathcal{L}_b]=0\ , \quad 
\partial_a \bar{\mathcal{L}}_b - \partial_b \bar{\mathcal{L}}_a +[\bar{\mathcal{L}}_a,\bar{\mathcal{L}}_b]=0\ . 
\label{Lflatness}
\end{equation}

The associated monodromy matrices are given by  
\begin{align}
 M(\tau,\sigma_2,\sigma_1;\lambda)&=P \exp\left[-\int^{\sigma_2}_{\sigma_1} d\sigma\,  \mathcal{L}_\sigma(\bm{\sigma};\lambda)\right]\ ,\\
\bar{M}(\tau,\sigma_2,\sigma_1;\lambda)&=P \exp\left[-\int^{\sigma_2}_{\sigma_1} d\sigma\,  \bar{\mathcal{L}}_\sigma(\bm{\sigma};\lambda)\right]\ ,
\end{align}
where $P$ is the path ordering. 
Here, we will only focus on $M(\tau,\sigma_2,\sigma_1;\lambda)$ and construct an infinite number of conserved charges. The same argument holds for $\bar{M}(\tau,\sigma_2,\sigma_1;\lambda)$ as well.
In the following, we will omit the arguments $\tau$ and $\lambda$ as $M(\tau,\sigma_2,\sigma_1;\lambda)=M(\sigma_2,\sigma_1)$ and $\mathcal{L}_a(\bm{\sigma};\lambda)=\mathcal{L}_a(\sigma)$ if there is no confusion.
Some basic properties of the monodromy matrix are summarized as below 
\begin{align}
 M(\sigma,\sigma)&=1\ ,\\
M(\sigma_3,\sigma_2)M(\sigma_2,\sigma_1)&=M(\sigma_3,\sigma_1)\ ,\\
M(\sigma_2,\sigma_1)^{-1}&=M(\sigma_1,\sigma_2)\ \\
\partial_{\sigma_2}M(\sigma_2,\sigma_1)&=-\mathcal{L}_\sigma(\sigma_2)M(\sigma_2,\sigma_1)\ ,\label{dM1}\\
\partial_{\sigma_1}M(\sigma_2,\sigma_1)&=M(\sigma_2,\sigma_1) \mathcal{L}_\sigma(\sigma_1)\ ,\label{dM2}\\
\delta M(\sigma_2,\sigma_1)&=-\int^{\sigma_2}_{\sigma_1} d\sigma\, M(\sigma_2,\sigma) \delta \mathcal{L}_\sigma(\sigma) M(\sigma,\sigma_1)\ ,\label{deltaM}
\end{align}
where $\delta$ denotes the variation with respect to $\mathcal{L}_\sigma(\sigma)$.

By using Eq.(\ref{deltaM}), the time derivative of the monodromy matrix is evaluated as follows:
\begin{equation}
 \partial_\tau M(\sigma_2,\sigma_1) = -\int^{\sigma_2}_{\sigma_1} d\sigma M(\sigma_2,\sigma) \partial_\tau \mathcal{L}_\sigma(\sigma) M(\sigma,\sigma_1)\ .
 \label{timederivativeM}
\end{equation}
Also, the flatness condition~(\ref{Lflatness}) leads to $\partial_\tau \mathcal{L}_\sigma=\partial_\sigma \mathcal{L}_\tau-[\mathcal{L}_\tau,\mathcal{L}_\sigma]$. 
Then, from Eqs.\,(\ref{dM1}) and (\ref{dM2}), we obtain 
\begin{align}
    \mathcal{L}_\sigma(\sigma) M(\sigma,\sigma_1)&= - \partial_\sigma M(\sigma,\sigma_1)\ , \nonumber \\
    M(\sigma_2,\sigma) \mathcal{L}_\sigma(\sigma) &= \partial_\sigma M(\sigma_2,\sigma)\ . \nonumber
\end{align}
As a result, the integrand of Eq.\,(\ref{timederivativeM}) becomes the total derivative with respect to $\sigma$. Thus, we obtain 
\begin{equation}
 \partial_\tau M(\sigma_2,\sigma_1) = M(\sigma_2,\sigma_1) \mathcal{L}_\tau(\sigma_1)-\mathcal{L}_\tau(\sigma_2) M(\sigma_2,\sigma_1)\ .
\label{dtauM}
\end{equation}

Let us suppose here a closed string with $0\leq \sigma \leq 2\pi$ and periodic boundary condition. 
From the above equation and the periodicity  $\mathcal{L}_a(2\pi)=\mathcal{L}_a(0)$, we find
\begin{equation}
 \partial_\tau \textrm{Tr} M_c(\tau,\lambda) = 0\ ,
\end{equation}
where $M_c(\tau,\lambda)\equiv M(\tau,2\pi,0;\lambda)$. 
We now have the conserved quantity as the one-parameter family of $\lambda$. The coefficients in the Taylor expansion of $\textrm{Tr}M_c$ 
give rise to an infinite number of conserved charges.

\subsection{Sufficient conditions for integrability of the open string}
\label{integrablecond}

In the above argument for the closed string, the periodicity $\sigma\sim \sigma +2\pi$ is essential. To apply this construction of an infinite number of conserved charges to the open string case, we need to take account of boundaries on the string worldsheet. In fact, under some special boundary conditions, an infinite number of conserved charges can be constructed even for open strings as well \cite{MacKay:2001bh,Delius:2001he,Mann:2006rh,Dekel:2011ja,MacKay:2011zs}.
For an open string, we will take the domain of the coordinate as $0\leq \sigma \leq \pi$ hereafter. 

Sufficient conditions for the integrability of the open string are given by the conditions for the boundary values of $J_\pm$ and $I_\pm$. 
The open string in AdS$_3$ is integrable if either (a), (b) or (c) is satisfied: 
\begin{equation}
\begin{split}
 &\textrm{(a)} \quad J_+^A=R^A{}_B J_-^B \ ,\\
&\textrm{(b)} \quad I_+^A=R^A{}_B I_-^B \ ,\\
&\textrm{(c)} \quad J_+^A=R^A{}_B I_-^B\textrm{ and }J_-^A=R^A{}_B I_+^B\ ,
\end{split}
\label{JIbc}
\end{equation}
where $J_\pm$ and $I_\pm$ are evaluated at the boundaries.
Here, $R^A{}_B$ is a constant matrix satisfying the following conditions:
\begin{itemize}
   \item $R$ is symmetric: $R^A{}_B= R_B{}^A$.
   \item $R$ is orthogonal: $R^A{}_C R_B{}^C = \delta^A_B$.
   \item $R$ gives a group automorphism by $T'{}^A=R^A{}_BT^B$: $[T'{}^{A},T'{}^B]=f^{AB}{}_CT'{}^C$.
\end{itemize}
The conditions (a) and (b) are called chiral boundary conditions. The condition (c) is called the achiral boundary condition. 

For the following arguments, it is convenient to introduce a linear map $\alpha$ defined as 
\begin{equation}
\alpha(T^A) \equiv R^A{}_B T^B\ . 
\end{equation}
For example, $\alpha$ acts on the current $J_a$ as 
\begin{equation}
    \alpha(J_a)=\alpha(J_{aA}T^A)=J_{aA}R^A{}_B T^B\ .
\end{equation}
One can see that the linear map $\alpha$ is an involution ($\alpha^2=1$) and automorphism from the definitions of $R$. 
Then, the boundary conditions in (\ref{JIbc}) are rewritten as 
\begin{equation}
\begin{split}
  &  \mbox{(a)} \quad J_+=\alpha(J_-)\ , \\ 
  & \mbox{(b)} \quad I_+=\alpha(I_-)\ ,  \\  
  & \mbox{(c)} \quad J_\pm = \alpha (I_\mp)\ .  
\end{split}
\end{equation}
Using the Lax pair~(\ref{Laxdef}), these conditions are also written as 
\begin{equation}
\begin{split}
&  \mbox{(a)} \quad \mathcal{L}_+(\bm{\sigma};\lambda)=\alpha(\mathcal{L}_-(\bm{\sigma};-\lambda))\ ,  \\  
&  \mbox{(b)} \quad \bar{\mathcal{L}}_+(\bm{\sigma};\lambda)=\alpha(\bar{\mathcal{L}}_-(\bm{\sigma};-\lambda))\ , \\ 
&  \mbox{(c)} \quad\mathcal{L}_\pm(\bm{\sigma};\lambda)=\alpha(\bar{\mathcal{L}}_\mp(\bm{\sigma};-\lambda))   
\end{split}
\end{equation}
for any values of $\lambda$.
In terms of $(\tau,\sigma)$-coordinates, they are written as 
\begin{equation}
\begin{split}
 &\textrm{(a) } \quad \mathcal{L}_\tau(\bm{\sigma};\lambda)=\alpha(\mathcal{L}_\tau(\bm{\sigma};-\lambda)), \quad \mathcal{L}_\sigma(\bm{\sigma};\lambda)=-\alpha(\mathcal{L}_\sigma(\bm{\sigma};-\lambda)), \\
&\textrm{(b) } \quad \bar{\mathcal{L}}_\tau(\bm{\sigma};\lambda)=\alpha(\bar{\mathcal{L}}_\tau(\bm{\sigma};-\lambda)), \quad \bar{\mathcal{L}}_\sigma(\bm{\sigma};\lambda)=-\alpha(\bar{\mathcal{L}}_\sigma(\bm{\sigma};-\lambda)),\\
&\textrm{(c) } \quad  \mathcal{L}_\tau(\bm{\sigma};\lambda)=\alpha(\bar{\mathcal{L}}_\tau(\bm{\sigma};-\lambda)), \quad \mathcal{L}_\sigma(\bm{\sigma};\lambda)=-\alpha(\bar{\mathcal{L}}_\sigma(\bm{\sigma};-\lambda)).
\end{split}
\label{Lbc}
\end{equation}
Again, the quantities in the above conditions are evaluated at the boundaries. 

Note that $\alpha(\mathcal{L}_a)$ and $\alpha(\bar{\mathcal{L}}_a)$ also satisfy the flatness conditions since $\alpha$ is an automorphism. This enables us to define monodromy matrices as
\begin{equation}
\begin{split}
M'(\tau,\sigma_2,\sigma_1;\lambda)&=P \exp\left[-\int^{\sigma_2}_{\sigma_1} d\sigma\, \alpha(\mathcal{L}_\sigma(\bm{\sigma};\lambda))\right]\ ,\\
\bar{M}'(\tau,\sigma_2,\sigma_1;\lambda)&=P \exp\left[-\int^{\sigma_2}_{\sigma_1} d\sigma\, \alpha(\bar{\mathcal{L}}_\sigma(\bm{\sigma};\lambda))\right]\ .
\end{split}
\end{equation}
For the conditions (a), (b) and (c), we define $M_c(\tau,\lambda)$ as
\begin{equation}
\begin{split}
&\textrm{(a) } \quad M_c(\tau,\lambda)=M'(\tau,0,\pi;-\lambda)M(\tau,\pi,0;\lambda)\ ,\\
&\textrm{(b) } \quad M_c(\tau,\lambda)=\bar{M}'(\tau,0,\pi;-\lambda)\bar{M}(\tau,\pi,0;\lambda)\ ,\\
&\textrm{(c) } \quad M_c(\tau,\lambda)=\bar{M}'(\tau,0,\pi;-\lambda)M(\tau,\pi,0;\lambda)\ ,
\end{split}
\end{equation}
Let us consider the time derivative of $M_c$ under the condition (a). From Eq.\ (\ref{dtauM}), it can be evaluated as 
\begin{equation}
\begin{split}
\partial_\tau M_c(\tau,\lambda)=&\big\{M'(\tau,0,\pi;-\lambda) \, \alpha\big(\mathcal{L}_\tau(\tau,\sigma=\pi;-\lambda)\big)\\
&-\alpha\big(\mathcal{L}_\tau(\tau,\sigma=0;-\lambda)\big) \, M'(\tau,0,\pi;-\lambda) \big\}\, M(\tau,\pi,0;\lambda)\\
&+M'(\tau,0,\pi;-\lambda)\big\{M(\tau,\pi,0;\lambda) \, \mathcal{L}_\tau(\tau,\sigma=0;\lambda)\\
&-\mathcal{L}_\tau(\tau,\sigma=\pi;\lambda) \, M(\tau,\pi,0;\lambda)\big\}\ .
\end{split}
\end{equation}
Then the condition (a) in Eq.(\ref{Lbc}) leads to $\partial_\tau \textrm{Tr}M_c=0$. Similarly, one can prove the conservation of $\textrm{Tr}M_c$ for conditions (b) and (c) as well.
The coefficients in the Taylor expansion of $\textrm{Tr}M_c$ give rise to an infinite number of conserved charges.

\section{Integrable boundary conditions for open strings in AdS$_3$}
\label{integrableAdSstring}

In this section, we shall classify possible boundary conditions preserving the integrability of an open string in AdS$_3$.

\begin{figure}[t]
\centering
\includegraphics[scale=0.6]{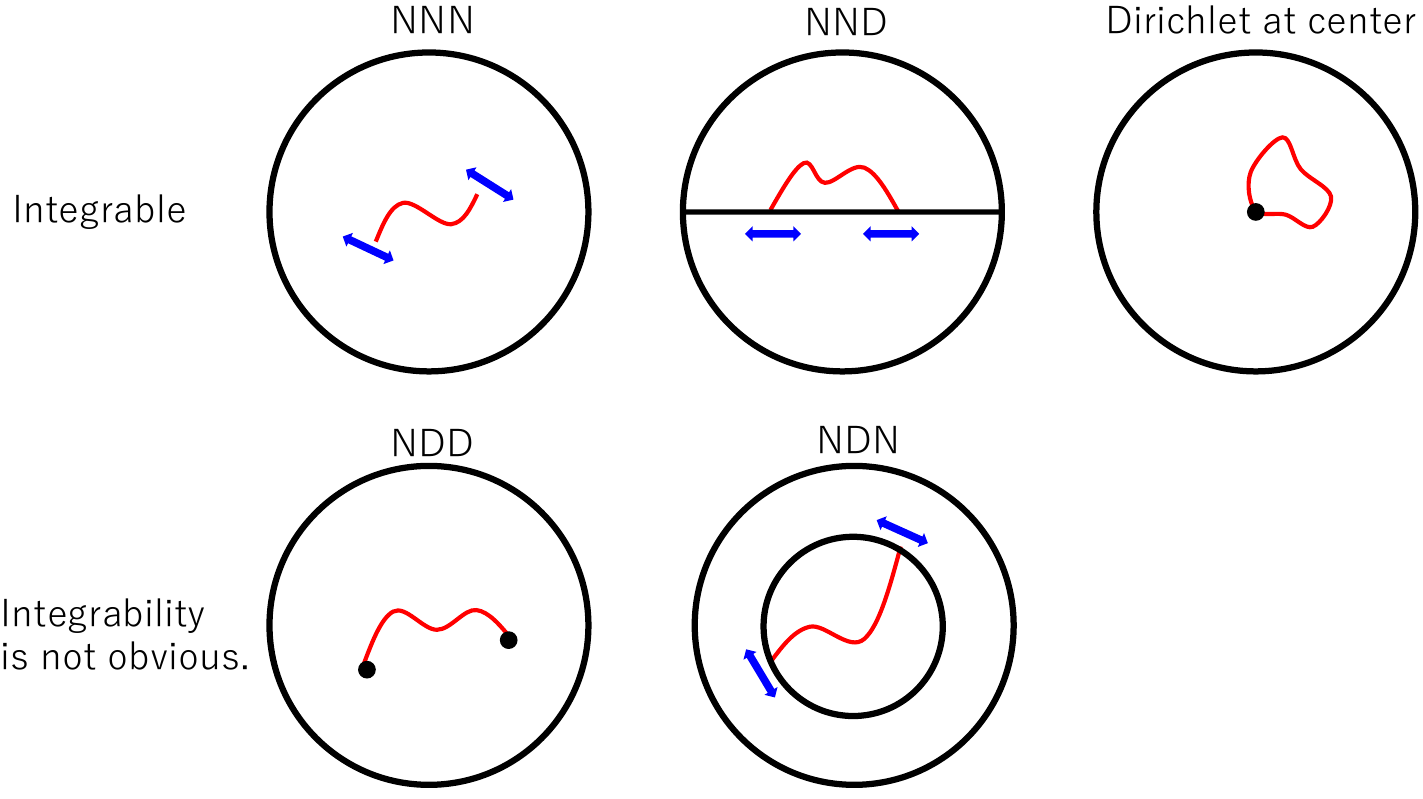}
\caption{Schematic profiles of an open string for NND, NND, NDD and NDN. The upper right is a special case of NDD, where both endpoints of the open string is fixed at the center of AdS$_3$.
}
\label{fig:string_ponchi}
\end{figure}

Let us consider a string near the $\sigma=0$ endpoint located in the AdS bulk ($r<1$). 
The target space coordinates in the global AdS$_3$ \eqref{dsglobal} are expanded as 
\begin{equation}
\begin{split}
t(\tau,\sigma)&=t_0(\tau) + t_1(\tau)\sigma + \cdots\ ,\\ 
r(\tau,\sigma)&=r_0(\tau) + r_1(\tau)\sigma + \cdots\ ,\\
\theta(\tau,\sigma)&=\theta_0(\tau)+\theta_1(\tau)\sigma + \cdots\ .
\end{split}
\label{generalbc}
\end{equation}
We will impose Neumann or Dirichlet boundary conditions on $t(\tau,\sigma),r(\tau,\sigma),\theta(\tau,\sigma)$.
For example, for $t(\tau,\sigma)$, 
the  Neumann and Dirichlet boundary conditions are described as $t_1(\tau)=0$ and $\dot{t}_0(\tau)=0$, respectively.
Either Neumann or Dirichlet boundary condition is imposed on each of the three variables. Therefore, there are $2^3=8$ choices. 
We shall denote them as NNN, NND, NDN and so on. This notation indicates the Neumann (N) or Dirichlet (D) boundary condition for $\{t,r,\theta\}$ in this order.
For example, NDD corresponds to $t_1=\dot{r}_0=\dot{\theta}_0=0$. In Fig.\,\ref{fig:string_ponchi}, schematic profiles of an open string are shown for several boundary conditions. Note that the case of Dirichlet boundary conditions for $r=0$ (see the upper-right figure in Fig.\ref{fig:string_ponchi}) has to be considered separately in the polar coordinates $(r,\theta)$. 

By using (\ref{generalbc}) with (\ref{Xglobal}), (\ref{J1}) and (\ref{IJandA}),
we find the boundary values of the invariant 1-forms at $\sigma=0$ as
\begin{equation}
 \begin{split}
J_+^0&=\alpha (\dot{t}_0+t_1) - \beta (\dot{\theta}_0+\theta_1)\ ,\\
J_+^1+iJ_+^2&=e^{-i(t_0 + \theta_0)}\{-\gamma(\dot{t}_0+t_1 - \dot{\theta}_0-\theta_1) + i \delta (\dot{r}_0 + r_1)\}\ ,\\
I_+^0&=-\alpha (\dot{t}_0+t_1) - \beta (\dot{\theta}_0+\theta_1)\ ,\\
I_+^1+iI_+^2&=ie^{i(t_0 - \theta_0)}\{\gamma(\dot{t}_0+t_1 + \dot{\theta}_0+\theta_1) + i \delta (\dot{r}_0 + r_1)\}\ ,
 \end{split}
\label{JpIpbdry}
\end{equation}
and
\begin{equation}
 \begin{split}
J_-^0&=\alpha (\dot{t}_0-t_1) - \beta (\dot{\theta}_0-\theta_1)\ ,\\
J_-^1+iJ_-^2&=e^{-i(t_0 + \theta_0)}\{-\gamma(\dot{t}_0-t_1 - \dot{\theta}_0+\theta_1) + i \delta (\dot{r}_0 - r_1)\}\ ,\\
I_-^0&=-\alpha (\dot{t}_0-t_1) - \beta (\dot{\theta}_0-\theta_1)\ ,\\
I_-^1+iI_-^2&=ie^{i(t_0 - \theta_0)}\{\gamma(\dot{t}_0-t_1 + \dot{\theta}_0-\theta_1) + i \delta (\dot{r}_0 - r_1)\}\ ,
 \end{split}
\label{JmImbdry}
\end{equation}
where 
\begin{equation}
 \alpha=\frac{(1+r_0^2)^2}{(1-r_0^2)^2}\ ,\quad \beta = \frac{4r_0^2}{(1-r_0^2)^2}\ ,\quad \gamma=\frac{2r_0(1+r_0^2)}{(1-r_0^2)^2}\ ,\quad \delta=\frac{2}{1-r_0^2}\ .
\end{equation}
From the double null constraints \eqref{virasoro} at $\sigma=0$ in the $(\tau,\sigma)$ coordinates \eqref{tausigma}, we obtain
\begin{align}
 -\frac{(1+r_0^2)^2}{4}\dot{t}_0 t_1 + \dot{r}_0 r_1 + r_0^2  \dot{\theta}_0 \theta_1&=0\ ,\\
 -\frac{(1+r_0^2)^2}{4}(\dot{t}_0^2+t_1^2) + \dot{r}_0^2+ r_1^2 + r_0^2 ( \dot{\theta}_0^2+ \theta_1^2)&=0\ . \label{4.6}
\end{align}
The first equation is automatically satisfied for the Neumann or Dirichlet boundary conditions on the $(t,r,\theta)$ coordinates. The second one introduces a relation among the three nonzero coefficients under the Neumann or Dirichlet boundary conditions. 

As an example, let us examine the integrability of NNN. Substituting $t_1=r_1=\theta_1=0$ into Eqs.(\ref{JpIpbdry}) and (\ref{JmImbdry}), we obtain the boundary values of the invariant 1-forms as
\begin{equation}
 \begin{split}
J_+^0&=\alpha \dot{t}_0 - \beta \dot{\theta}_0\ ,\quad 
J_+^1+iJ_+^2=e^{-i(t_0 + \theta_0)}\{-\gamma(\dot{t}_0 - \dot{\theta}_0) + i \delta \dot{r}_0\}\ ,\\
I_+^0&=-\alpha \dot{t}_0 - \beta \dot{\theta}_0\ ,\quad 
I_+^1+iI_+^2 =ie^{i(t_0 - \theta_0)}\{\gamma(\dot{t}_0 + \dot{\theta}_0) + i \delta \dot{r}_0\}\ ,
\end{split}
\end{equation}
and 
\begin{equation}
 \begin{split}
J_-^0&=\alpha \dot{t}_0 - \beta \dot{\theta}_0\ ,\quad
J_-^1+iJ_-^2 =e^{-i(t_0 + \theta_0)}\{-\gamma(\dot{t}_0 - \dot{\theta}_0) + i \delta \dot{r}_0\}\ ,\\
I_-^0&=-\alpha \dot{t}_0 - \beta \dot{\theta}_0\ ,\quad 
I_-^1+iI_-^2 =ie^{i(t_0 - \theta_0)}\{\gamma(\dot{t}_0 + \dot{\theta}_0) + i \delta \dot{r}_0\}\ .
 \end{split}
\end{equation}
From these expressions, we find linear relations between the $+$ and $-$ sectors as
\begin{equation}
 J_+^A=J_-^A\ ,\quad I_+^A=I_-^A\ \qquad (A=0,1,2)\ .
 \label{JpmIpm_NNN_relation}
\end{equation}
The constant matrix $R$ defined in Eq.(\ref{JIbc}) is $R^A{}_B=\delta^A_B$. 
This is manifestly symmetric and orthogonal. Also $T'{}^A=R^A{}_B T^B$ gives an automorphism. Therefore, NNN is integrable.

As another example, let us consider the NDD case. Substituting $t_1=\dot{r}_0=\dot{\theta}_0=0$ into Eqs.(\ref{JpIpbdry}) and (\ref{JmImbdry}), we have 
\begin{equation}
 \begin{split}
J_+^0&=\alpha \dot{t}_0 - \beta \theta_1\ ,\quad
J_+^1+iJ_+^2=e^{-it_0}\{-\gamma(\dot{t}_0 -\theta_1) + i \delta r_1\}\ ,\\
I_+^0&=-\alpha \dot{t}_0 - \beta \theta_1\ ,\quad
I_+^1+iI_+^2=ie^{it_0}\{\gamma(\dot{t}_0 +\theta_1) + i \delta r_1\}\ ,
\end{split}
\end{equation}
and 
\begin{equation}
 \begin{split}
J_-^0&=\alpha \dot{t}_0 + \beta \theta_1\ ,\quad 
J_-^1+iJ_-^2=e^{-it_0}\{-\gamma(\dot{t}_0 +\theta_1) - i \delta  r_1\}\ ,\\
I_-^0&=-\alpha \dot{t}_0 + \beta\theta_1\ ,\quad
I_-^1+iI_-^2=ie^{it_0}\{\gamma(\dot{t}_0 -\theta_1) - i \delta  r_1\}\ ,
 \end{split}
\end{equation}
where we set $\theta_0=0$ without loss of generality.
We cannot find any linear relations between the $+$ and $-$ sectors for $r_0\neq 0$. Hence, there is no indication of integrability.
(The case of $r_0=0$ will be treated later.)

Similar analysis is possible for each boundary condition. The results are summarized in Table~\ref{table:integrability}. 
(See Appendix~\ref{examineall} for detailed analysis.)
Under the boundary conditions marked ``True'', the open string in AdS$_3$ is integrable.
For the boundary conditions with ``?'', integrability is not guaranteed by this analysis (the conditions in section~\ref{integrablecond}).
In the following sections, we will study the dynamics of the string under NDD and show that it exhibits weak turbulence.
This gives evidence that the string under NDD boundary conditions is non-integrable.

\begin{table}[hbtp]
  \centering
  \begin{tabular}{|c|c|}
    \hline
    Boundary conditions for ($t,r,\theta$)  & Integrability \\
    \hline \hline
    NNN  & True \\
    NND  & True \\
    NDN  & ?  \\
    NDD  & ?  \\
    DNN  & ? \\
    DND  & True \\
    DDN  & ? \\
    DDD  & ? \\
    \hline
  \end{tabular}
\caption{Integrability for open strings with Neumann and Dirichlet boundary conditions.}
  \label{table:integrability}
\end{table}

Note that the case of $r|_{\sigma=0}=r_0=0$ needs to be treated separately.
This corresponds to the situation that the string endpoint is located at the center of AdS$_3$. (See the upper-right figure in Fig.\,\ref{fig:string_ponchi}.)
Since the center of AdS$_3$ is the coordinate singularity in the polar coordinates ($t,r,\theta$), 
we have to use the Cartesian coordinates (\ref{AdScartes}). 
In the Cartesian coordinates, $r|_{\sigma=0}=0$ corresponds to $x|_{\sigma=0}=y|_{\sigma=0}=0$.
Expanding the target space coordinates near the string endpoint as
$x(t,\sigma)=x_1(\tau)\sigma + x_2(\tau)\sigma^2+\cdots$ and $y(t,\sigma)=y_1(\tau)\sigma + y_2(\tau)\sigma^2+\cdots$, we have
\begin{equation}
 \begin{split}
J_+^0&= t_1 + \dot{t}_0\ ,\quad
J_+^1+iJ_+^2=2ie^{-it_0}(x_1-iy_1)\ ,\\
I_+^0&= -t_1 - \dot{t}_0\ ,\quad
I_+^1+iI_+^2=-2e^{it_0}(x_1-iy_1)\ ,\\
J_-^0&= -t_1 + \dot{t}_0\ ,\quad
J_-^1+iJ_-^2=-2ie^{-it_0}(x_1-iy_1)\ ,\\
I_-^0&= t_1 - \dot{t}_0\ ,\quad
I_-^1+iI_-^2=2e^{it_0}(x_1-iy_1)\ .
 \end{split}
\end{equation}
Imposing the Neumann boundary condition on $t(\tau,\sigma)$, i.e.~$t_1=0$, we find linear relations
\begin{equation}
 J_+^A=R^A{}_B J_-^B\ , \quad I_+^A=R^A{}_B I_-^B\ ,\quad  
R=\textrm{diag}(1,-1,-1)\ .
\end{equation}
Both satisfy the integrability condition in section~\ref{integrablecond}. 
Imposing the Dirichlet condition to $t(\tau,\sigma)$ instead, i.e.~$\dot{t}_0=0$, we find
\begin{equation}
 J_+^A=R^A{}_B J_-^B\ , \quad I_+^A=R^A{}_B I_-^B\ ,\quad  
R^A{}_B=-\delta^A_B\ .
\end{equation}
One can check that $T'{}^A=R^A{}_B T^B$ is not an automorphism. Therefore, the integrability condition is not satisfied in this case.
The result is summarized in Table~\ref{table:integrability2}.

\begin{table}[hbtp]
  \centering
  \begin{tabular}{|c|c|}
    \hline
    Boundary conditions for $t$  & Integrability \\
    \hline \hline
    N  & True \\
    D  & ? \\
    \hline
  \end{tabular}
\caption{Integrability in the case that the string endpoint is located at the center of AdS$_3$.}
  \label{table:integrability2}
\end{table}

Finally, it is worth to make a comment on the D-brane interpretation. In the preceding works \cite{Skenderis:2002vf,Sakaguchi:2003py,Sakaguchi:2004md},  possible supersymmetric AdS-branes are classified. In particular, 1/2 BPS AdS-branes are anticipated to provide integrable open strings. It may be rather natural because the 1/2 BPS AdS D-branes are static and so the associated integrability may be preserved as well. For example, the (2,0)-brane sitting at the origin of AdS in the work \cite{Sakaguchi:2003py} corresponds to the NND case. 
The NNN case may be seen as an AdS$_3$ subspace of the (3,1)-brane, (4,2)-brane or (5,3)-brane in \cite{Sakaguchi:2003py}. The result in Table\,\ref{table:integrability2} (the upper-right figure in Fig.\,\ref{fig:string_ponchi}) 
may be seen as a part of the (1,3)-brane case.

\section{Nonlinear perturbation of an open string attached to the AdS boundary}
\label{sec:nonlinear}

In the rest of this paper, we study numerically time evolution of the nonlinear oscillation of a classical Nambu-Goto open string hanging from the AdS boundary in the global AdS. In a limit, this setup contains the nonlinear waves on a straight string in AdS \cite{Mikhailov:2003er}. In the literature, however, the effect of wave reflection at the string endpoints (i.e.~worldsheet boundaries) was not fully incorporated. Here, we consider full nonlinear time evolution of the string including the boundaries.

The boundary conditions for the string studied in this and next sections would naturally correspond to NDD in the classification given in the previous section. This is because the spatial location of the string endpoints $(r,\theta)$ are fixed in late time in time evolution. Therefore, we expect that integrability is not guaranteed. However, when the endpoints are located on the AdS boundary, the asymptotic series expansion is different from \eqref{generalbc}, and the notion of N and D boundary conditions needs to be adjusted accordingly. We will comment on this later in this section.

\subsection{Static string}
\label{sec:static}

For initial configurations, we consider a static open string attached to the AdS boundary in global AdS$_3$. This corresponds to the holographic dual description of the static quark-antiquark potential in $\mathcal{N}=4$ super Yang-Mills theory on $R \times S^3$. By symmetry, the open string is embedded in $\mathrm{AdS}_3 \subset \mathrm{AdS}_5$. The same setup has been studied in \cite{BallonBayona:2008uc}. Here, in addition, we give the embedding function of the string explicitly.

To write the static string solution, we introduce the $z$ coordinate $( 0 < z \le 1)$ as
\begin{equation}
z=\frac{1-r^2}{1+r^2} \ .
\label{r2z_coord}
\end{equation}
The metric \eqref{dsglobal} is rewritten as
\begin{equation}
ds^2=\frac{1}{z^2} \left( -dt^2 + \frac{dz^2}{1-z^2} + (1-z^2) d\theta^2 \right) \ .
\end{equation}
The AdS boundary and the center of the AdS are located at $z=0$ and $z=1$, respectively. In a static gauge $(\tau,\sigma)=(t,z)$, the embedding of the string can be specified by a function $\theta(z)$. With this ansatz, the Nambu-Goto action \eqref{NambuGotoAction} becomes
\begin{equation}
S = -\frac{1}{2\pi\alpha'} \int dt dz \frac{1}{z^2} \sqrt{\frac{1 + (1-z^2)^2 \theta^{\prime}(z)^2}{1-z^2}} \ .
\label{gads_stat_NG}
\end{equation}

The embedding of the open string can be given by elliptic integrals. Without loss of generality, we can assume that the string is line symmetric in $\theta \to -\theta$ and smooth at $\theta=0$. Let $z=z_0 \ ( \le 1)$ denote the coordinate where $\theta(z_0)=0$ on the string. The string then extends in $0 < z \le z_0$. The string embedding equation is obtained from \eqref{gads_stat_NG} as
\begin{equation}
\theta'(z)= \mp \frac{z^2}{(1-z^2)\sqrt{z_0^4(1-z^2)/(1-z_0^2) - z^4}} \ ,
\label{gads_stat_embeddingeq}
\end{equation}
where the upper and lower signs are chosen for $\theta>0$ and $\theta<0$, respectively. Under the boundary condition $\theta(z_0)=0$, \eqref{gads_stat_embeddingeq} can be solved by
\begin{equation}
\theta(z) = \pm \frac{\sqrt{1-z_0^2}}{z_0} \left[ F\left( \frac{z}{z_0} ; i \sqrt{1-z_0^2} \right) - \Pi \left(z_0^2 ; \frac{z}{z_0} ; i \sqrt{1-z_0^2} \right) - F_0 + \Pi_0\right],
\label{global_static_qqbar_z_sol}
\end{equation}
where $F(x;k)$ and $\Pi(n;x;k)$ are incomplete elliptic integrals defined as
\begin{align}
F(x;k) &= \int^x_0 \frac{dt}{\sqrt{(1-t^2)(1-k^2 t^2)}} \ , \\
\Pi(n;x;k) &= \int^x_0 \frac{dt}{(1-n \hspace{1pt} t^2) \sqrt{(1-t^2)(1-k^2 t^2)}} \ ,
\end{align}
and $F_0$ and $\Pi_0$ are their complete elliptic integrals,
\begin{align}
F_0 \equiv  F \left(1;i \sqrt{1-z_0^2} \right) \ , \quad \Pi_0 \equiv \Pi \left(z_0^2 ; 1 ; i \sqrt{1-z_0^2} \right) \ .
\end{align}

\begin{figure}[t]
\centering
\centering
\subfigure[Static strings]{\includegraphics[scale=0.45]{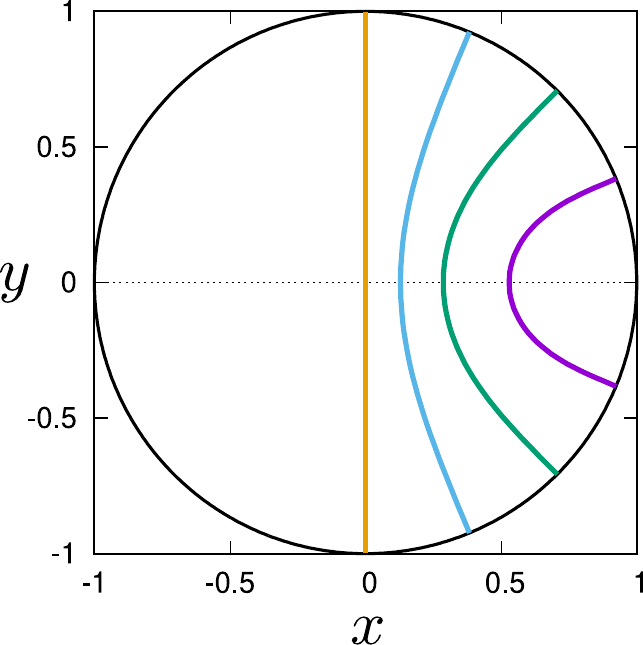}\label{fig:staticqqbar}}\qquad
\subfigure[$\varphi$ coordinate]{\includegraphics[scale=0.45]{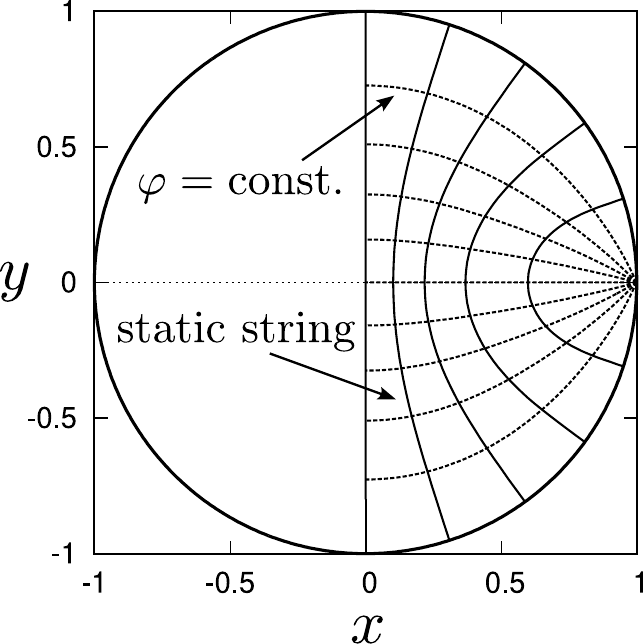}\label{fig:varphigrid}}
\caption{Left: The static strings for $\theta_b/(\pi/2)=0.25,\,0.5,\,0.75,\,1$. Right: $\varphi$ coordinate and static strings. Dotted lines denote constant $\varphi$ (for $\varphi=0.1,0.2,\dots,0.9$ from top to bottom, with $\varphi=0$ and 1 on the AdS boundary), and the real lines are the static strings (for $\theta_b/(\pi/2)=0.2,\,0.4,\,0.6,\,0.8,\,1$).}
\label{fig:global_statstring}
\end{figure}

The embedding can be specified by the location of the string endpoints on the boundary, denoted by $\theta = \pm \theta_b \ (0 \le \theta_b \le \pi/2)$. It is related to $z_0$ as
\begin{equation}
\theta_b = \lim_{z \to 0} |\theta(z)| = \frac{\sqrt{1-z_0^2}}{z_0} (-F_0+\Pi_0) \ .
\label{static_qqbar_psi_bdry}
\end{equation}
In figure~\ref{fig:staticqqbar}, the static strings are shown for $\theta_b/(\pi/2)=0.25,\,0.5,\,0.75,\,1$. The endpoint coordinate approaches $\theta_b \to 0$ and $\theta_b \to \pi/2$ in the limits $z_0 \to 0$ and $z_0 \to 1$, respectively. In $\theta_b \to 0$, the string shrinks to a point, where with an appropriate zooming in, the quark-antiquark open string in Poincar\'{e} AdS \cite{Rey:1998ik,Maldacena:1998im} can be obtained. When $\theta_b=\pi/2$, the string embedding is straight at $\theta(z)=\pi/2$, where the worldsheet is $\mathrm{AdS}_2 \subset \mathrm{AdS}_3$. We call this case as the {\it antipodal string}, connecting $\theta=\pm \pi/2$ of the boundary $S^1$.

To parametrize the static string, we find it convenient to use another coordinate. In $\theta >0$, we introduce a new coordinate $\phi \ (0 \le \phi < F_0$) by the inverse function of $F( z/z_0 ; i \sqrt{1-z_0^2} )$ as
\begin{equation}
\frac{z}{z_0} = \mathrm{sn}\left(\phi;i\sqrt{1-z_0^2}\right) \ ,
\label{z_sn_gauge}
\end{equation}
where $\phi=0$ and $\phi=F_0$ correspond to $z=0$ and $z=z_0$ on the string, respectively. The positive sign solution of \eqref{global_static_qqbar_z_sol} can be rewritten as
\begin{equation}
\theta(\phi) = \frac{\sqrt{1-z_0^2}}{z_0} \left[ \phi - \Pi \left(z_0^2 ; \mathrm{sn}\left( \phi;i\sqrt{1-z_0^2}\right) ; i \sqrt{1-z_0^2} \right) -F_0 +\Pi_0\right] \ .
\label{phi_sol_sn_gauge}
\end{equation}
In $\theta<0$, we set $F_0 < \phi \le 2 F_0$ with $\phi|_{z=0}=2F_0$. Using $\tilde{\phi} \equiv 2 F_0 - \phi$, we can write
\begin{equation}
\frac{z}{z_0} = \mathrm{sn}\left(\tilde{\phi};i\sqrt{1-z_0^2}\right) \ . \label{z_sn_gauge_2}
\end{equation}
The negative sign solution of \eqref{global_static_qqbar_z_sol} is then expressed as
\begin{equation}
\theta(\phi) = - \frac{\sqrt{1-z_0^2}}{z_0} \left[ \tilde{\phi} - \Pi \left(z_0^2 ; \mathrm{sn}\left( \tilde{\phi};i\sqrt{1-z_0^2}\right) ; i \sqrt{1-z_0^2} \right) -F_0 +\Pi_0\right] \ .
\label{phi_sol_sn_gauge_2}
\end{equation}
Note that $F_0$ depends on $z_0$. By introducing $\varphi \equiv \phi/(2 F_0)$, the parameter range can be fixed to $0 \le \varphi \le 1$. The $\varphi$ coordinate is shown in figure~\ref{fig:varphigrid}.

\subsection{Evolution equations}
We will study nonlinear time evolution when the string is perturbed. For numerical calculation, we use the ``Cartesian'' coordinates in \eqref{AdScartes} because they are regular at the center of AdS, $(x,y)=(0,0)$. Let $\bm{\chi}=(x,y)$ denote these coordinates collectively. In the worldsheet double null coordinates $(\sigma^+,\sigma^-)$, the induced metric is given by
\begin{align}
h_{++} &= \frac{1}{(1-|\bm{\chi}|^2)^2} \left( -(1+|\bm{\chi}|^2)^2 \, t_{,+}^2 + 4 |\bm{\chi}_{,+}|^2 \right) \ , \label{ind_hpp}\\
h_{--} &= \frac{1}{(1-|\bm{\chi}|^2)^2} \left( -(1+|\bm{\chi}|^2)^2 \, t_{,-}^2 + 4 |\bm{\chi}_{,-}|^2 \right) \ , \label{ind_hmm} \\
h_{+-} &= \frac{1}{(1-|\bm{\chi}|^2)^2} \left( -(1+|\bm{\chi}|^2)^2 \, t_{,+} \, t_{,-} + 4 \bm{\chi}_{,+} \cdot \bm{\chi}_{,-} \right) \ , \label{ind_hpm}
\end{align}
where $|\bm{\chi}|^2=r^2=x^2+y^2$. The first two should satisfy the constraint equations $h_{\pm\pm}=0$ \eqref{virasoro}. In these coordinates, the Nambu-Goto action is given by
\begin{align}
S &= \frac{1}{2\pi\alpha'}\int d\sigma^+d\sigma^- h_{+-} \nonumber \\
&= \frac{1}{2 \pi \alpha'} \int d\sigma^+d\sigma^- \frac{1}{(1-|\bm{\chi}|^2)^2} \left( -(1+|\bm{\chi}|^2)^2 \, t_{,+} \, t_{,-} + 4 \bm{\chi}_{,+} \cdot \bm{\chi}_{,-} \right) \ .
\end{align}
The equations of motion are
\begin{align}
t_{,+-} &= -\frac{4}{1-|\bm{\chi}|^4} \bm{\chi} \cdot \left(t_{,+} \, \bm{\chi}_{,-} + \bm{\chi}_{,+} \, t_{,-} \right) \ , \label{eomuv_t} \\
\bm{\chi}_{,+-} &= -\frac{1}{1-|\bm{\chi}|^2} \Big( \bm{\chi}(1+|\bm{\chi}|^2) t_{,+} \, t_{,-} \nonumber \\
&\hspace*{14ex} + 2 \big(\bm{\chi}_{,+} (\bm{\chi} \cdot \bm{\chi}_{,-}) + (\bm{\chi}_{,+} \cdot \bm{\chi}) \bm{\chi}_{,-} - (\bm{\chi}_{,+} \cdot \bm{\chi}_{,-})\bm{\chi} \big) \Big) \ .
\label{eomuv_chi}
\end{align}

If these equations are used as they are, however, time evolution is unstable. This is because the relation of the directions of the target space and worldsheet times is not specified. Solving the constraint equations $h_{\pm\pm}=0$, we can fix the sign of $t_{,+}$ and $t_{,-}$ as
\begin{align}
t_{,+} = \frac{2 |\bm{\chi}_{,+}|}{1+|\bm{\chi}|^2} >0 \ , \quad
t_{,-} = \frac{2 |\bm{\chi}_{,-}|}{1+|\bm{\chi}|^2} >0\ .
\label{constraintsuv_chi}
\end{align}
These relations ensure that $t_{,+}>0$ and $t_{,-}>0$ are future directed vectors. Applying \eqref{constraintsuv_chi} to (\ref{eomuv_t}) and (\ref{eomuv_chi}), we can realize stable time evolution.

\subsection{Asymptotic solutions near the AdS boundary}

We comment on the asymptotic series near the AdS boundary and integrable boundary conditions when the string endpoints is located at the AdS boundary.

When the equations of motion \eqref{eomuv_t} and \eqref{eomuv_chi} are solved near the AdS boundary $r=1$, the form of the asymptotic solutions is different from \eqref{generalbc}. Near the $\sigma=0$ endpoint, we find
\begin{align}
t(\tau,\sigma) &= t_0(\tau) + \frac{(\dot{t}_0^2+\dot{\theta}_0^2)\ddot{t}_0-2\dot{t}_0\dot{\theta}_0\ddot{\theta}_0}{2(\dot{t}_0^2-\dot{\theta}_0^2)}\sigma^2 + t_3(\tau)\sigma^3 + \cdots \ , \\
r(\tau,\sigma) &= 1 - (\dot{t}_0^2-\dot{\theta}_0^2)^{1/2} \sigma + \frac{\dot{t}_0^2-\dot{\theta}_0^2}{2}\sigma^2 + r_3(\tau)\sigma^3 + \cdots \ , \\
\theta(\tau,\sigma) &= \theta_0(\tau) + \frac{2\dot{t}_0\dot{\theta}_0\ddot{t}_0-(\dot{t}_0^2+\dot{\theta}_0^2)\ddot{\theta}_0}{2(\dot{t}_0^2-\dot{\theta}_0^2)}\sigma^2 + \theta_3(\tau)\sigma^3 + \cdots \ ,
\end{align}
where we assume $\dot{t}_0\neq 0$, and $t_3$ and $r_3$ are fixed by solving the constraint equations \eqref{constraintsuv_chi} as 
\begin{align}
t_3(\tau) &= \frac{\dot{\theta}_0}{\dot{t}_0}\theta_3 \ , \\
r_3(\tau) &= \frac{\dot{\theta}_0\dddot{\theta_0}-\dot{t}_0\dddot{t_0}}{6(\dot{t}_0^2-\dot{\theta}_0^2)^{1/2}} +  \frac{(\dot{\theta}_0\ddot{t}_0-\dot{t}_0\ddot{\theta}_0)^2}{2(\dot{t}_0^2-\dot{\theta}_0^2)^{3/2}} + \frac{(\dot{t}_0^2-\dot{\theta}_0^2)^{1/2}}{6}(\dot{\theta}_0^2-2\dot{t}_0^2) \ .
\end{align}
In the asymptotic expansion, $t_0, \theta_0$ and $\theta_3$ are not determined.

When the above asymptotic expansions are used as they are, we find that the currents diverge as $I^A_\pm, J^A_\pm \sim 1/\sigma^2$. This implies that it will be necessary to remove the divergence by renormalizing the currents in order to define finite currents. Even without doing that treatment, nevertheless, we can still get some insights. Taking a linear combination of the bare currents, we find
\begin{align}
J^0_+ - J^0_- &= -\frac{2(\ddot{t}_0-\ddot{\theta}_0)}{(\dot{t}_0-\dot{\theta}_0)^2\sigma} - \frac{6 \theta_3}{\dot{t}_0(\dot{t}_0-\dot{\theta}_0)} + O(\sigma) \ , \\
\frac{(J^1_+ + iJ^2_+) - (J^1_- + iJ^2_-)}{e^{i(t_0-\theta_0)} } &=\left(2+\frac{2i(\ddot{t}_0-\ddot{\theta}_0)}{(\dot{t}_0-\dot{\theta}_0)^2} \right) \frac{1}{\sigma} + \frac{6 i \theta_3}{\dot{t}_0(\dot{t}_0-\dot{\theta}_0)} + O(\sigma) \ , \\
\frac{(J^1_+ - iJ^2_+) - (J^1_- - iJ^2_-)}{e^{-i(t_0-\theta_0)} } &=\left(2-\frac{2i(\ddot{t}_0-\ddot{\theta}_0)}{(\dot{t}_0-\dot{\theta}_0)^2} \right) \frac{1}{\sigma} - \frac{6 i \theta_3}{\dot{t}_0(\dot{t}_0-\dot{\theta}_0)} + O(\sigma) \ , \\
I^0_+ - I^0_- &= \frac{2(\ddot{t}_0+\ddot{\theta}_0)}{(\dot{t}_0+\dot{\theta}_0)^2\sigma} - \frac{6 \theta_3}{\dot{t}_0(\dot{t}_0+\dot{\theta}_0)} + O(\sigma) \ , \\
\frac{(I^1_+ + iI^2_+) - (I^1_- + iI^2_-)}{e^{-i(t_0+\theta_0)} } &=\left(-2i-\frac{2(\ddot{t}_0+\ddot{\theta}_0)}{(\dot{t}_0+\dot{\theta}_0)^2} \right) \frac{1}{\sigma} + \frac{6 \theta_3}{\dot{t}_0(\dot{t}_0+\dot{\theta}_0)} + O(\sigma) \ , \\
\frac{(I^1_+ - iI^2_+) - (I^1_- - iI^2_-)}{e^{i(t_0+\theta_0)} } &=\left(2i-\frac{2(\ddot{t}_0+\ddot{\theta}_0)}{(\dot{t}_0+\dot{\theta}_0)^2} \right) \frac{1}{\sigma} + \frac{6 \theta_3}{\dot{t}_0(\dot{t}_0+\dot{\theta}_0)} + O(\sigma) \ .
\end{align}
If $\theta_3=0$, the finite part satisfies the same linear relations as \eqref{JpmIpm_NNN_relation}. Therefore, integrability would be guaranteed for $\theta_3=0$. For a static straight antipodal string, we have $\theta_3=0$, but otherwise we have $\theta_3 \neq 0$ in general (i.e.~non-antipodal static strings and dynamically perturbed strings). If $\theta_3 \neq 0$, however, we do not find any linear relations between the $''+''$ and $''-''$ sectors. Hence, integrability would not be guaranteed. For time evolution of the string with $\theta_3 \neq 0$, we do not expect integrability. We are interested in the time evolution under $\dot{\theta}_0=0$. Then, the situation would be analogous to NDD.

\subsection{Initial data and boundary conditions for time evolution}

To calculate the time evolution for \eqref{eomuv_t}--\eqref{constraintsuv_chi}, we need initial data and boundary conditions of the string.

For the initial data, we parametrize by worldsheet coordinates the static string solution given in terms of target space coordinates in section~\ref{sec:static}. In terms of $(\tau,\sigma)$ \eqref{tausigma}, the sum of the constraint equations \eqref{constraintsuv_chi} is written as
\begin{equation}
\dot{t}^2 = \frac{4(r'^2 + r^2 \theta'^2)}{(1+r^2)^2} \ .
\label{constraint_tdotsquare}
\end{equation}
where $\dot{} \equiv \partial_\tau{}$ and ${}' \equiv \partial_\sigma$. We choose the spatial domain of the open string worldsheet as $0 \le \sigma \le \pi$ (i.e.~$\sigma=\pi \varphi$). Then, this equation can be solved by
\begin{equation}
t = \frac{2 F_0 z_0}{\pi} \tau = \frac{2 F_0 z_0}{\pi}(\sigma^++\sigma^-) \ , \quad 
\phi =\frac{2 F_0}{\pi} \sigma = \frac{2 F_0}{\pi}(\sigma^+-\sigma^-) \ ,
\label{mygauge}
\end{equation}
where \eqref{tausigma} is used. We substitute this parametrization to \eqref{z_sn_gauge}--\eqref{phi_sol_sn_gauge_2}. Furthermore, the $(z,\theta)$ coordinates are converted to $(x,y)$ by using \eqref{xy_coordinates} and \eqref{r2z_coord}. Then, we can prepare the initial data $(t(\sigma^+,\sigma^-),x(\sigma^+,\sigma^-),y(\sigma^+,\sigma^-))$.
As a special case, when the string embedding is antipodal $\theta(\sigma)=\pi/2$, the static string can be simply expressed as
\begin{equation}
t= \tau \ , \quad x=0 \ , \quad 
y = \tan\left( \frac{\pi}{4} - \frac{\sigma}{2} \right) \ .
\end{equation}

For the boundary conditions, we move one endpoint of the string for a short duration of time $\Delta t$. We fix the endpoint at $\sigma=\pi$ to $\theta=-\theta_b$ and move the other end at $\sigma=0$ around $\theta=\theta_b$. We use the following profile for the endpoint motion:
\begin{equation}
\theta(t) = \theta_b + \epsilon \alpha(t) \ ,
\label{quench_longi}
\end{equation}
where $\epsilon$ specifies the amplitude and $\alpha(t)$ is chosen as a compactly supported $C^\infty$ function given by
\begin{equation}
\alpha(t)=
\begin{cases}
\exp\left[ 2\left(\frac{\Delta t}{t-\Delta t}-\frac{\Delta t}{t}+4\right) \right]\qquad &(0<t<\Delta t)\\
0\qquad &(\textrm{otherwise})
\end{cases} \ .
\end{equation}
The endpoint is brought back to the original location $\theta=\theta_b$ after the time $\Delta t$.

We solve the time evolution of the string \eqref{eomuv_t}--\eqref{constraintsuv_chi} under the aforementioned initial data and boundary condition. In numerical calculations, we discretize the equations of motion in the double null coordinates by second-order finite difference as utilized in \cite{Ishii:2014paa,Ishii:2015wua}. Our numerical calculations are second-order convergent as $(\Delta h)^2$ with respect to the grid size of the finite difference $\Delta h$. Many of our results were obtained with discretization of $2^{11}=2048$ segments in the computational domain of $\sigma^+$ and $\sigma^-$ (i.e.~$\Delta h = \pi/2^{11}$). We also carried out calculations with different number of segments and checked numerical convergence.

\subsection{Cusp formation}

As a result of the time evolution, cusps can be formed on the hanging string in AdS$_3$ \cite{Ishii:2015wua}. Such cusps can be detected by evaluating the Jacobian of the map from the worldsheet to target space. For the string worldsheet in the three dimensional target space, we have three Jacobians $J_{tx}, J_{ty},J_{xy}$. On top of a cusp, all of them vanish simultaneously as
\begin{equation}
J_{ij} = \chi_{i,+} \chi_{j,-} - \chi_{i,-} \chi_{j,+}=0 \ ,
\label{jacobian_J}
\end{equation}
where $\chi_i=t,x,y$ $(i=0,1,2)$. Numerically, we identify the presence of a cusp if all $J_{ij}$ change signs simultaneously in a discretized numerical grid.

\subsection{Linear perturbation and energy spectrum}

To evaluate nonlinearity, we expand the nonlinearly oscillating string by linear eigenmodes around the static string and see deviations.

To this end, we first calculate the linear perturbation of the static string. Here, we express the static solution by $\bar{x}(\varphi)$, $\bar{y}(\varphi)$.\footnote{Here, $\bar{x},\bar{y},\bar{\bm{\chi}}$ are used in a different meaning from the coordinates $(\bar{t},\bar{r},\bar{\theta})$ in \eqref{Xtwist}, but the distinction would be clear.} In the $(x,y)$-plane, a unit vector normal to the static string can be given by $(\hat{n}_x,\hat{n}_y)=(\bar{y}'/|\bar{\bm{\chi}}'|,-\bar{x}'/|\bar{\bm{\chi}}'|)$, where $'\equiv\partial_\varphi$. Then, linear fluctuation of the static string can be introduced as
\begin{equation}
x(t,\varphi) = \bar{x}(\varphi) + \hat{n}_x \xi (t,\varphi) \ , \quad
y(t,\varphi) = \bar{y}(\varphi) + \hat{n}_y \xi (t,\varphi) \ .
\label{lin_flucansatz}
\end{equation}
The quadratic action for $\xi$ takes the form
\begin{equation}
S = \frac{1}{2 \pi\alpha'} \int dt d\varphi \frac{1}{2} \left( C_t \dot{\xi}^2 -  C_\varphi \xi'^2 - V \xi^2 \right) \ ,
\label{lin_quadaction}
\end{equation}
where $\dot{}\equiv\partial_t$, and
\begin{equation}
C_t = \frac{8z_0 F_0}{(1-\bar{x}^2-\bar{y}^2)^2} \ , \quad
C_\varphi = \frac{2}{z_0 F_0 (1-\bar{x}^2-\bar{y}^2)^2} \ ,
\label{lin_quad_coeffs}
\end{equation}
while $V$ is lengthy and we do not reproduce it here. The linearized equation of motion for the fluctuation is given by
\begin{equation}
(\partial_t^2 + \mathcal{H}) \xi = 0 \ , \quad 
\mathcal{H} \equiv - \frac{1}{C_t} \partial_\varphi C_\varphi \partial_\varphi + \frac{V}{C_t} \ .
\label{lin_quad_eom}
\end{equation}
The operator $\mathcal{H}$ is Hermitian under the following inner product:
\begin{equation}
\langle \alpha,\beta \rangle \equiv \int_0^1 d\varphi \, C_t(\varphi) \alpha(\varphi) \beta(\varphi) \ .
\end{equation}

\begin{figure}[t]
\centering
\subfigure[Eigenfrequencies]{\includegraphics[scale=0.45]{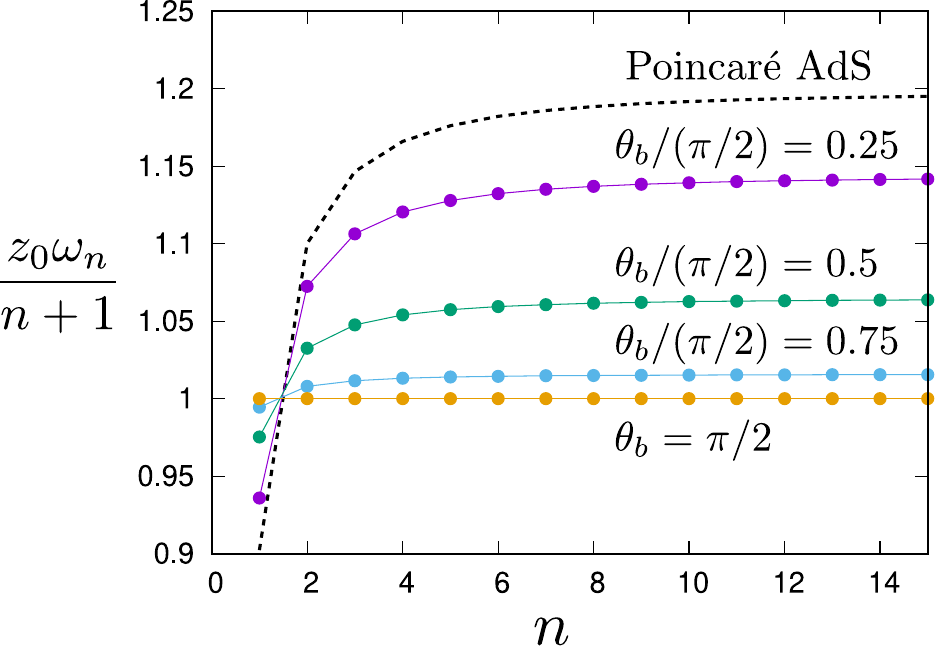}\label{fig:eigenfreqL}}\qquad
\subfigure[Eigenfunctions for $\theta_b=\pi/2$]{\includegraphics[scale=0.45]{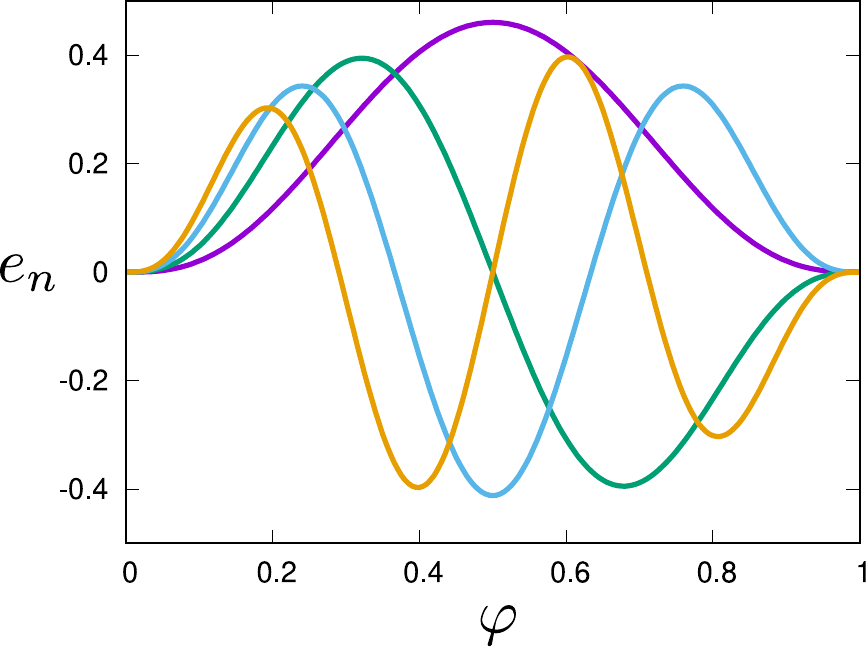}\label{fig:eigenfnAP}}
\caption{Left: Eigenfrequencies for $\theta_b/(\pi/2)=0.25,\,0.5,\,0.75,\,1$. The data points for the same $\theta_b$ are connected by lines for visibility. The black dashed line corresponds to the shrinking string limit $\theta_b \to 0$, where the eigenfrequencies match those in Poincar\'{e} AdS \cite{Ishii:2015wua}. Right: The first four $(n=1,2,3,4)$ eigenfunctions for the antipodal string \eqref{lin_ap_eigenfn}.}
\label{fig:eigenfnAPs}
\end{figure}

We obtain the eigenfrequencies and eigenfunctions $(\omega_n, e_n) \ (n=1,2,3\cdots)$ as solutions to the eigenvalue problem \eqref{lin_quad_eom}. The eigenfunctions are normalized as $\langle e_m,e_n \rangle=\delta_{mn}$. When the static string is not antipodal, we need to solve the eigenvalue problem numerically. For the antipodal string, the eigenvalue problem can be analytically solved by $\omega_n = n+1$ and 
\begin{equation}
e_n(\varphi) = -\sqrt{ \frac{4\zeta^2+\omega_n^2(1-\zeta^2)^2}{2\pi (\omega_n^2-1)} } \cos \left( \frac{\omega_n \pi}{2} - 2\omega_n \tan^{-1} \! \zeta + \tan^{-1} \! \frac{2\zeta}{\omega_n(1-\zeta^2)} \right) \ ,
\label{lin_ap_eigenfn}
\end{equation}
where
\begin{equation}
\zeta \equiv
\begin{cases}
\hfill \sqrt{(1-\sin (\pi \varphi))/(1+\sin (\pi \varphi))} & (0 \le \varphi \le 0.5) \\
-\sqrt{(1-\sin (\pi \varphi))/(1+\sin (\pi \varphi))} & (0.5 \le \varphi \le 1)
\end{cases} \ .
\end{equation}
In figure~\ref{fig:eigenfreqL}, the eigenfrequencies are shown for $\theta_b/(\pi/2)=0.25,\,0.5,\,0.75,\,1$. In the limit $\theta_b \to 0$, the frequencies match those in Poincar\'{e} AdS \cite{Ishii:2015wua}, denoted by the black dashed line. In figure~\ref{fig:eigenfnAP}, the first four eigenfunctions of the antipodal string \eqref{lin_ap_eigenfn} are shown.

We then expand nonlinear string solutions by the linear eigenmodes. In the same manner as \eqref{lin_flucansatz}, we define nonlinear fluctuation $\Xi(t,\varphi)$ by
\begin{equation}
x = \bar{x}(\varphi) + \hat{n}_x \Xi(t,\varphi) \ , \quad
y = \bar{y}(\varphi) + \hat{n}_y \Xi(t,\varphi) \ .
\label{nlin_flucdef}
\end{equation}
When the amplitude of the nonlinear fluctuation is small, we decompose $\Xi$ in terms of the linear eigenmodes as
\begin{equation}
\Xi(t,\varphi) = \sum_{n=1}^{\infty} c_n(t) e_n(\varphi) \ .
\label{nlin_modedecomp}
\end{equation}
Using the above decomposition in the quadratic action \eqref{lin_quadaction} where $\xi$ is replaced with $\Xi$, we can construct the energy spectrum of the dynamical string as
\begin{equation}
\varepsilon_n(t) = \frac{1}{4\pi \alpha'} \left( \dot{c}_n^2 + \omega_n^2 c_n^2 \right) \ .
\end{equation}
We will use this to evaluate the growth of nonlinearity.

\section{Results}
\label{sec:results}

In this section, we discuss numerical results of the time evolution of the string. 

\subsection{Non-antipodal string}
\label{sec:nonantipodal}

\begin{figure}[t]
\centering
\subfigure[early times]{\includegraphics[scale=0.45]{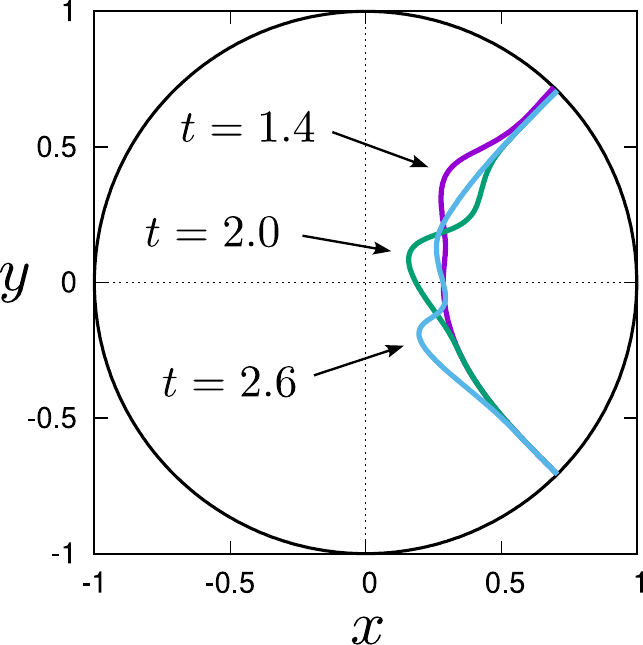}\label{fig:snapshot_pi4a}}\qquad
\subfigure[after cusp formation]{\includegraphics[scale=0.45]{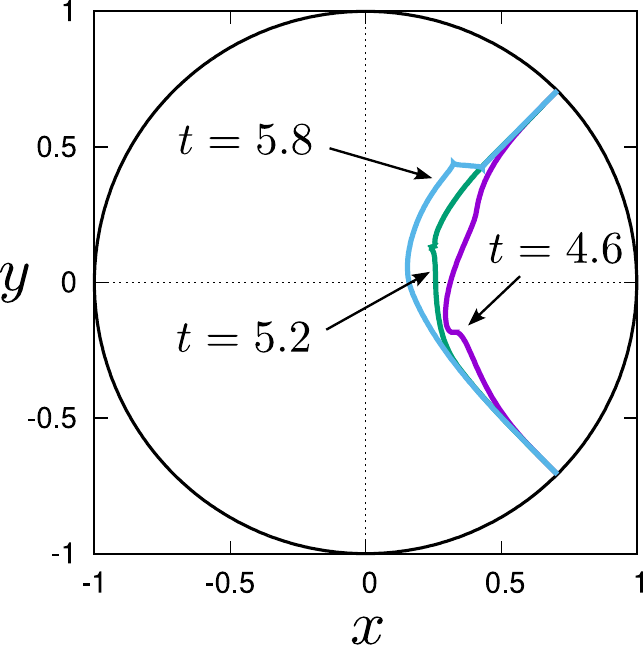}\label{fig:snapshot_pi4b}}
\caption{Snapshots for $\theta_b=\pi/4, \, \epsilon=0.08, \, \Delta t=2$.}
\label{fig:snapshot_pi4}
\end{figure}

We first examine the case that the string endpoints are away from the antipodal points. In figure~\ref{fig:snapshot_pi4}, string snapshots are shown for $\theta_b=\pi/4, \, \epsilon=0.08, \, \Delta t=2$. The perturbation is introduced from the upper end of the string for $0 \le t \le 2$. The wave induced on the string is initially smooth as seen in figure~\ref{fig:snapshot_pi4a}. It gets sharper as it propagates on the string, and a cusp pair is formed around $t \simeq 4.4$. The cusps can be visually recognized in later times as shown in figure~\ref{fig:snapshot_pi4b}. The pair-created cusps separate and remain propagating on the string. This is the same as the case of cusps on the string in Poincar\'{e} AdS \cite{Ishii:2015wua}.

\begin{figure}[t]
\centering
\includegraphics[scale=0.45]{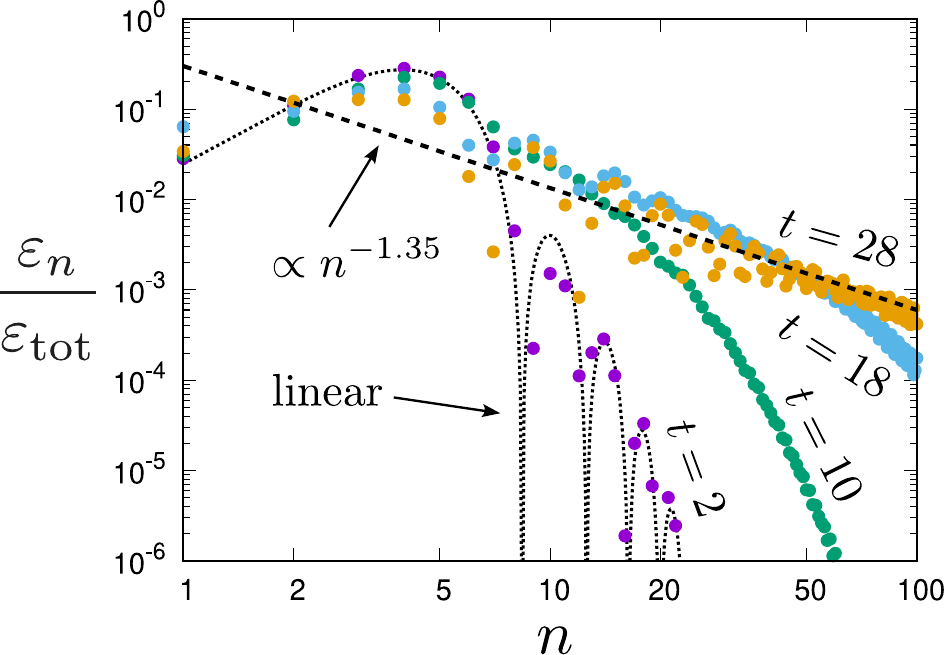}
\caption{Time dependence of the energy spectrum for $\theta_b=\pi/4, \, \epsilon=0.01, \, \Delta t=2$. In the plot, $\varepsilon_n$ is normalized by the total energy of the linear energy spectrum $\varepsilon_\mathrm{tot}$. The black dotted curve is an interpolation of the linear energy spectrum for the linearized action \eqref{lin_quadaction}. The dashed line is a fit of $t=28$ energy spectrum given by a power law $\epsilon_n \propto n^{-1.35}$.}
\label{fig:ensp_pi4_001}
\end{figure}

The behavior toward the cusp formation indicates turbulence on the string, characterized by the transfer of energy from large to small scales. In figure~\ref{fig:ensp_pi4_001}, the energy spectra are shown for $\theta_b=\pi/4, \, \epsilon=0.01, \, \Delta t=2$. In the black dotted curve (labeled ``linear''), for reference, we also show the linear energy spectrum for the linearized action \eqref{lin_quadaction}, where the spectrum is calculated by using the linear response theory under the boundary perturbation \eqref{quench_longi} in the same way as it was done in \cite{Ishii:2015wua}. Note that the energy spectrum is obtained for integer $n$, but here it is shown in an interpolated curve for visibility. The linear energy spectrum is independent of time because nonlinear terms are ignored. This implies that the time dependent difference of data points from this black dotted curve indicates nonlinearity. In the figure, the mode energy $\varepsilon_n$ is normalized by the total energy of the linear energy spectrum $\varepsilon_\mathrm{tot}$. Because the amplitude of the perturbation is small as $\epsilon=0.01$, the energy spectrum just after introducing the boundary perturbation is quite similar to the linear spectrum ($t=2$: purple dots). The energy spectrum then changes in time ($t=10,18,28$).

In figure \ref{fig:ensp_pi4_001}, the energy spectrum saturates a power law dependence toward the cusp formation in late times. This manifests turbulence -- transfer of energy from large to small scales because of nonlinearity. For this figure, cusps are formed at $t_\mathrm{cusp} \simeq 28.2$. The energy spectrum is calculated before this cusp formation time because the eigenmode expansion is ill-defined if cusps exist on the string. A fit of the spectrum just before the cusp formation is approximately $\epsilon_n \propto n^{-1.35}$.

\begin{figure}[t]
\centering
\includegraphics[scale=0.45]{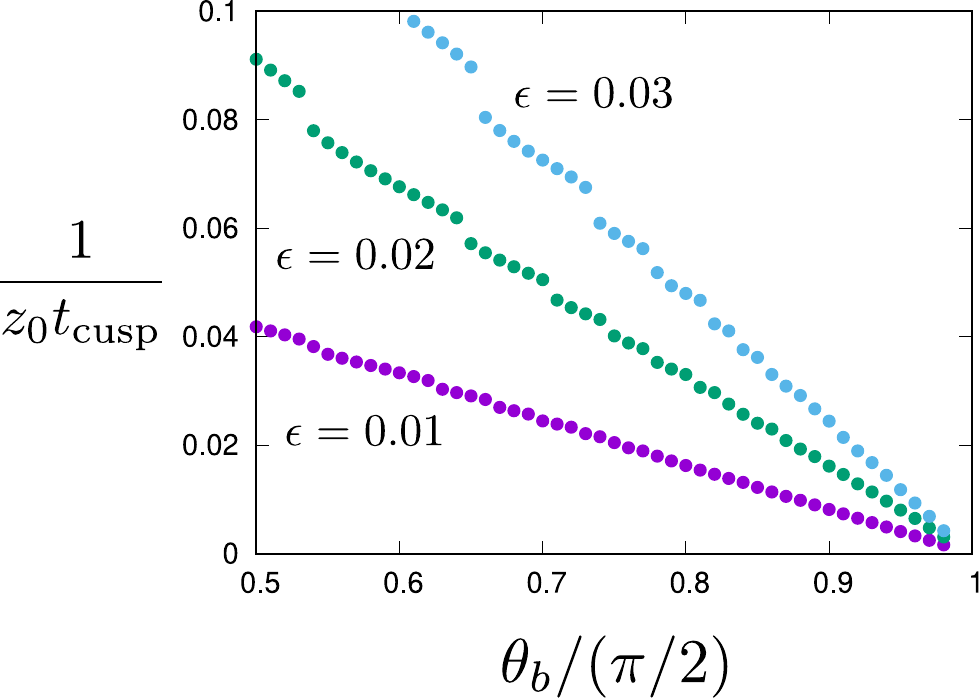}
\caption{Cusp formation time for non-antipodal strings for $\Delta t=2$. Data points are calculated for $\theta_b/(\pi/2) \le 0.98$. The data points almost overlap if $1/(\epsilon z_0 t_\mathrm{cusp})$ is plotted.}
\label{fig:tcusp_psi}
\end{figure}

Changing the locations of the endpoints affects cusp formation time. In figure~\ref{fig:tcusp_psi}, the time for the first cusp formation is plotted when $\theta_b$ and $\epsilon$ are varied while $\Delta t$ is fixed. The data points are given for $\theta_b/(\pi/2) \le 0.98$ and $\epsilon=0.01,0.02,0.03$. As the string embedding approaches antipodal, it takes longer before cusps to form. When $\pi/2-\theta_b$ is finite, the behavior can be fitted by
\begin{equation}
t_\mathrm{cusp} \simeq \frac{1}{5.19z_0 \epsilon (\pi/2-\theta_b)}.
\label{tcusp_fit}
\end{equation}
However, corrections will be significant to this fit in the antipodal limit $\theta_b \to \pi/2$. In particular, we wonder if $t_\mathrm{cusp}$ diverges or not in this limit. In the following, we directly study the nonlinear perturbations of the antipodal string.

\subsection{Antipodal string}
\label{sec:antipodal}

\begin{figure}[t]
\centering
\subfigure[focusing of waves]{\includegraphics[scale=0.45]{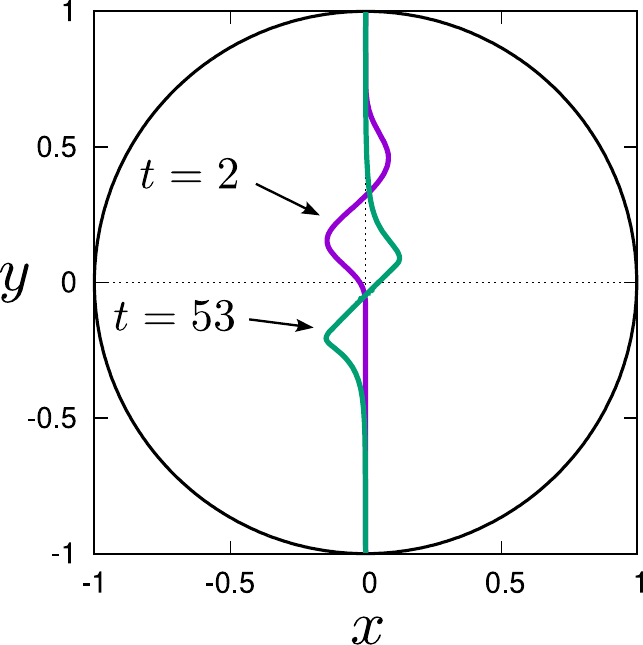}\label{fig:snapshot_pi2a}}\qquad
\subfigure[defocusing of waves]{\includegraphics[scale=0.45]{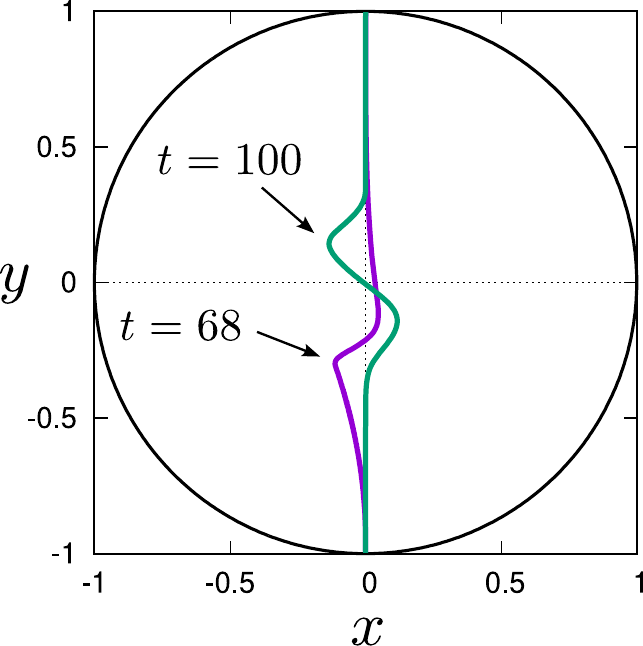}\label{fig:snapshot_pi2b}}
\caption{Snapshots for $\theta_b=\pi/2, \, \epsilon=0.1, \, \Delta t=2$.}
\label{fig:snapshot_pi2}
\end{figure}

Let us consider the case that the string endpoints are located at the antipodal points. In figure~\ref{fig:snapshot_pi2}, string snapshots for $\theta_b=\pi/2, \, \epsilon=0.1, \, \Delta t=2$ are shown. We find that the waves on the string get sharper initially (figure~\ref{fig:snapshot_pi2a}), but the string get loosened afterward (figure~\ref{fig:snapshot_pi2b}). Then, the waves get sharp again. With the parameters for the figure, this cycle repeats around an interval of  $t \sim 100$, but the recurrence is not completely periodic. The waves on the string gradually get sharper as the string experiences the cycles of focusing and defocusing. Continuing the time evolution, eventually we find that cusps are indeed detected on the string around $t \sim 700$ for these parameters.

\begin{figure}[t]
\centering
\subfigure[$\Delta t=2$]{\includegraphics[scale=0.45]{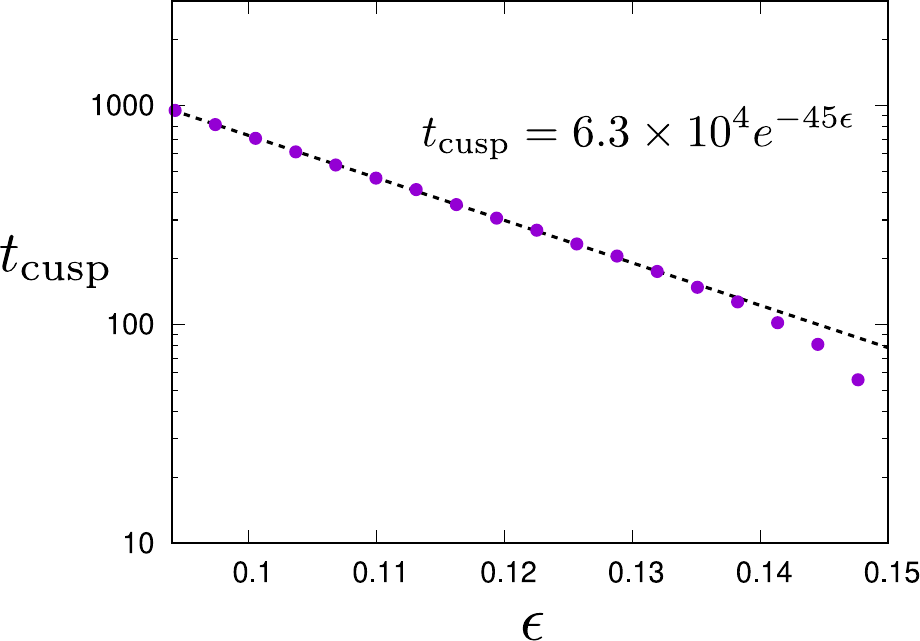}\label{fig:tcusp_ap_dt2}}\qquad
\subfigure[$\Delta t=8$]{\includegraphics[scale=0.45]{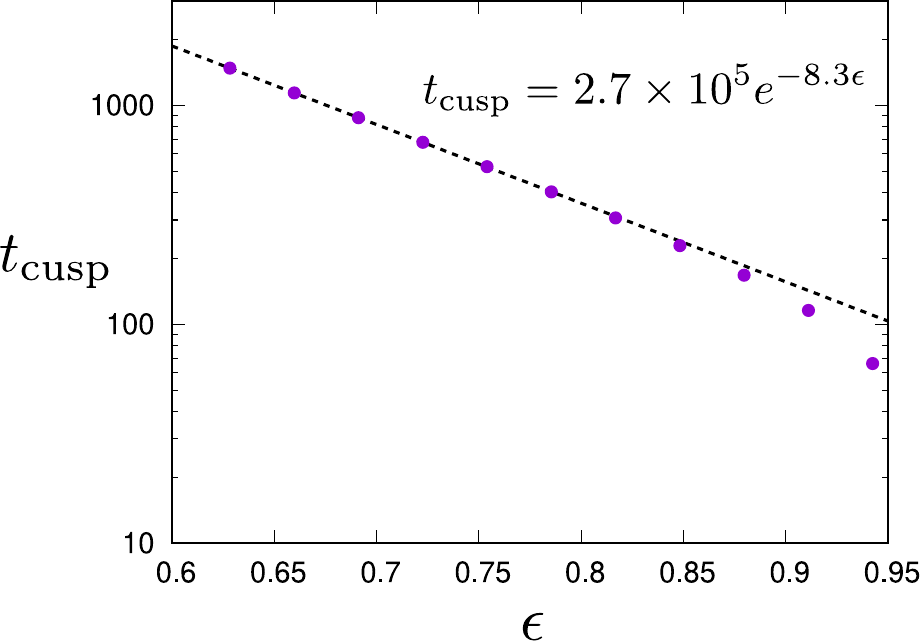}\label{fig:tcusp_ap_dt8}}
\caption{First cusp formation time for the antipodal string when $\epsilon$ is varied.}
\label{fig:tcusp_ap}
\end{figure}

We would like to add comments on this ``first detection'' of cusp formation for the antipodal string. When $\theta_b$ of non-antipodal strings is changed closer to the antipodal limit $\theta_b \to \pi/2$, the main contribution for cusp formation is switched from the collapse of waves to apparently collisions of sharpening waves. The pair-creation of cusps by the overlapping of nonlinear waves is momentary, and such cusps pair-annihilate immediately. Hence, we find that the first detection of cusps does not necessarily result in persistent cusps on the string. The cusp creation and annihilation then repeats along with the time evolution, with the lifetime of the cusp pairs getting longer. Eventually, cusp pairs separate and start to propagate on the string, but we need to wait much longer before this happens as the amplitude $\epsilon$ is decreased. For this reason, we focus on the first cusp creation time, which will be followed by the appearance of separating cusps in later times.

In figure~\ref{fig:tcusp_ap}, we plot the first cusp formation time for the nonlinearly perturbed antipodal string when $\epsilon$ is varied. We show results for $\Delta t=2$ and 8. The downward bent in large $\epsilon$ is due to the large amplitude of the perturbation, and we are not interested in that part. We focus on small $\epsilon$ instead. The cusp formation time gets exponentially longer as $\epsilon$ is decreased, but it appears that the antipodal string can form cusps for finite $\epsilon$. From these results, we expect turbulence (i.e.~growth of nonlinearity) for finite $\epsilon$, which we will discuss below.

\begin{figure}[t]
\centering
\subfigure[$\epsilon=0.01$]{\includegraphics[scale=0.45]{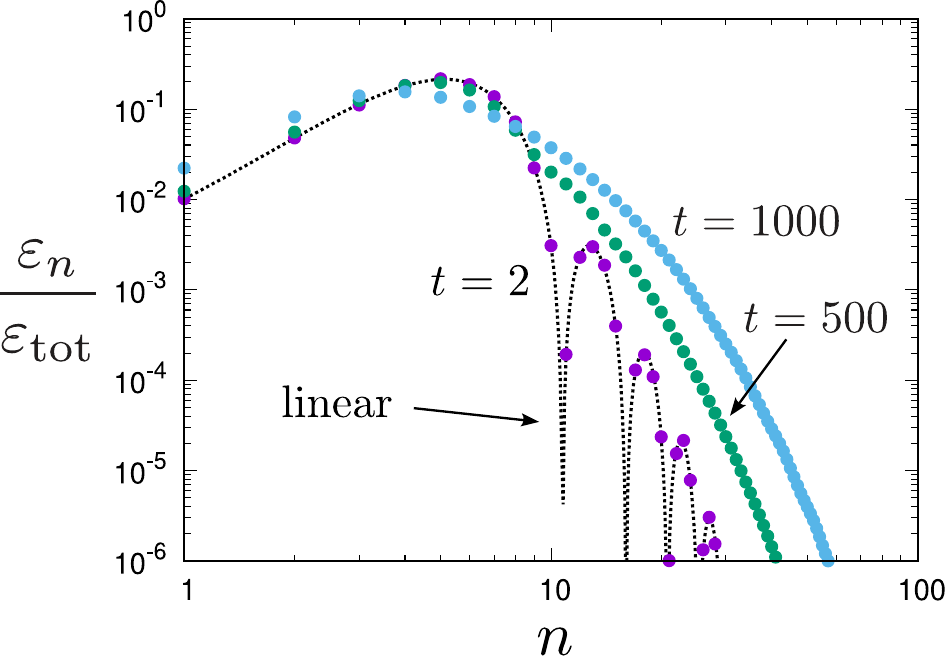}\label{fig:ensp_pi2_001}}\quad
\subfigure[$\epsilon=0.1$]{\includegraphics[scale=0.45]{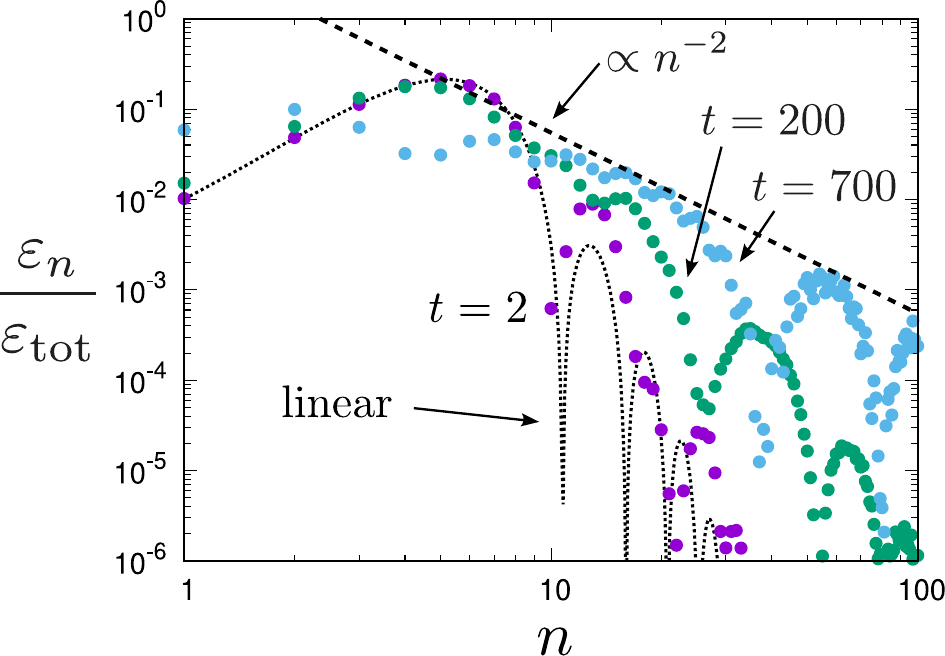}\label{fig:ensp_pi2_01}}
\caption{The time dependence of the energy spectrum for $\theta_b=\pi/2, \, \Delta t=2$ and $\epsilon=0.01$ (left), 0.1 (right). The black dotted curve is an interpolated function for the linear energy spectrum calculated from the linearized action \eqref{lin_quadaction}, and $\varepsilon_n$ is normalized by the total energy of the linear energy spectrum $\varepsilon_\mathrm{tot}$.}
\label{fig:ensp_pi2}
\end{figure}

In figure~\ref{fig:ensp_pi2}, the energy spectra for the perturbed antipodal string with $\Delta t=2$ are shown for $\epsilon=0.01$ and 0.1. In the case of the antipodal string, the transfer of the energy to higher modes is slower than non-antipodal cases. Therefore, it will take long for the energy spectrum to saturate a power law, but we can observe the presence of steady turbulent behavior in the energy spectrum even in early times. In figure~\ref{fig:ensp_pi2_001}, the initial energy spectrum (at $t=2$) is close to the linear theory. Then, the energy spectrum grows in time. For this figure ($\epsilon=0.01$), we did not continue the time evolution until cusps are formed, but apparently the energy spectrum will reach a power law. In figure~\ref{fig:ensp_pi2_01}, we used a larger amplitude $\epsilon=0.1$. The energy spectrum at the early time $t=2$ is distributed similarly to the linear one. At $t=700$, the energy spectrum is power law distributed. The power law in fact has been saturated already around $t \sim 400$. The envelope decays as $n^{-2}$. In summary, we observe the turbulence in the energy spectrum of a nonlinear oscillating open string even when the endpoints are located at the antipodal points on the AdS boundary.

\section{Conclusion}
\label{sec:conc}

In this paper, we have investigated the origin of the weak turbulence on the worldsheet of an open string in AdS. The classical dynamics of the string in AdS can be described by that of the PCM, which without boundaries is known to be integrable. We consider open strings. They correspond to the PCMs with boundaries, that can break integrability depending on the types of boundary conditions. We argued that integrability breaking at endpoints of the open string is the origin of turbulence. In the first part of this paper, we classified boundary conditions that preserve integrability or do not guarantee it. The results are summarized in Tables~\ref{table:integrability} and \ref{table:integrability2}. For example, when the endpoints of the open string are fixed in AdS (corresponding to NDD in our classification), integrability is not guaranteed. It would be likely that integrability is broken for such boundary conditions and turbulence would occur on the open string. In the second part of this paper, we have numerically investigated the nonlinear dynamics of the classical open string for the aforementioned boundary condition. We considered the open string hanging from the AdS boundary and calculated nonlinear time evolution of the string, and we observed weak turbulence. In particular, the weak turbulence was recognized even for an antipodal string. This means that, while nonlinear wave solutions were obtained for a straight string \cite{Mikhailov:2003er}, they are subject to the weak turbulence if the boundary conditions of the string worldsheet are taken into account.

In this paper, turbulence was numerically investigated for boundary conditions where integrability was not guaranteed. However, there is no proof that turbulence only occurs in such cases, although these cases are speculated to be non-integrable. In particular, among the patterns listed in Tables~\ref{table:integrability} and \ref{table:integrability2}, we tested only the NDD case for an open string that has endpoints on the AdS boundary. It will be an important future work to investigate other boundary conditions and clarify the differences in the classical dynamics of the open strings for integrable and non-integrable boundary conditions \cite{Kitaku}.

While the source of non-integrability can be introduced by the boundary conditions on open string endpoints, it is not evident whether the effect of the boundary conditions is localized on the endpoints or leaks to the bulk. As seen in figure~\ref{fig:snapshot_pi4b}, the collapse of nonlinear waves and formation of cusps can occur in the middle of the string, suggesting that the waves could experience focusing effect during their propagation on the string, not only at the moment of reflection on the endpoints. This observation is accompanied by the power law in the energy spectrum in Figure~\ref{fig:ensp_pi4_001}, which picks up the tendency of the growth of higher modes on the string and might not be only of the endpoint effects. It would be interesting to consider how the effect of the boundary conditions at the endpoint is transmitted to the bulk part of the string. Related argument has been given in \cite{Vegh:2018dda} that the boundary perturbation for the open string connected to the AdS boundary introduces a source of cusp formation propagating on the string worldsheet. It will also be interesting to generalize the viewpoint of (non-)integrable boundary conditions to integrable models with non-integrable boundary conditions in general.

\acknowledgments
The authors are grateful to Dimitrios Giataganas, Koji Hashimoto, Ryo Kitaku and Chulmoon Yoo for useful discussions and would also like to thank the hospitality of OIST during the workshop ``Integrability, Deformations and Chaos''. The discussion there was very helpful to finish this work. 
The work of T.I. was supported in part by JSPS KAKENHI Grant Number 19K03871.
The work of K.M. was supported in part by JSPS KAKENHI Grant Nos. 20K03976, 21H05186 and 22H01217. 
The work of K.Y. was supported by MEXT KAKENHI Grant-in-Aid for Transformative Research Areas A ``Extreme Universe'' No.~22H05259 and ``Machine Learning Physics'' No.~22H05115, and JSPS Grant-in-Aid for Scientific Research (B) No.~22H01217. 

\appendix

\section{Detailed analysis of integrable conditions}
\label{examineall}

In this appendix, we will complete the analysis of the integrable boundary conditions of the open string in AdS$_3$. 
The cases of NNN and NDD have been studied in section~\ref{integrableAdSstring}.
Here, we assume $r_0\neq 0$ when we impose the Dirichlet condition to $r(\tau,\sigma)$;
the analysis of the exceptional $r_0=0$ case has been done in section~\ref{integrableAdSstring}.

\subsection*{NND}
For $t_1=r_1=\dot{\theta}_0=0$, Eqs.(\ref{JpIpbdry}) and (\ref{JmImbdry}) become
\begin{equation}
 \begin{split}
J_+^0&=\alpha \dot{t}_0 - \beta \theta_1\ ,\quad J_+^1+iJ_+^2=e^{-i t_0}\{-\gamma(\dot{t}_0 -\theta_1) + i \delta\dot{r}_0\}\ ,\\
I_+^0&=-\alpha \dot{t}_0 - \beta \theta_1\ ,\quad I_+^1+iI_+^2=ie^{it_0}\{\gamma(\dot{t}_0 +\theta_1) + i \delta \dot{r}_0\}\ ,\\
J_-^0&=\alpha \dot{t}_0 + \beta \theta_1\ ,\quad J_-^1+iJ_-^2=e^{-i t_0}\{-\gamma(\dot{t}_0+\theta_1) + i \delta \dot{r}_0\}\ ,\\
I_-^0&=-\alpha \dot{t}_0 + \beta\theta_1\ ,\quad I_-^1+iI_-^2=ie^{i t_0}\{\gamma(\dot{t}_0 -\theta_1) + i \delta \dot{r}_0 \}\ .
 \end{split}
\end{equation}
We find linear relations, 
\begin{equation}
J_+^A=R^A{}_B I_-^B\ ,\quad I_+^A=R^A{}_B J_-^B\ ,\quad 
R=
\begin{pmatrix}
-1 & 0 & 0 \\
0 & 0 & -1 \\
0 & -1 & 0
\end{pmatrix}
\ .
\end{equation}
$R$ is symmetric and orthogonal. Also $T'{}^A=R^A{}_B T^B$ gives an automorphism. Therefore, NND is integrable.

\subsection*{NDN}
For $t_1=\dot{r}_0=\theta_1=0$, Eqs.(\ref{JpIpbdry}) and (\ref{JmImbdry}) become
\begin{equation}
 \begin{split}
J_+^0&=\alpha \dot{t}_0 - \beta \dot{\theta}_0\ ,\quad
J_+^1+iJ_+^2=e^{-i(t_0 + \theta_0)}\{-\gamma(\dot{t}_0 - \dot{\theta}_0) + i \delta r_1)\}\ ,\\
I_+^0&=-\alpha \dot{t}_0 - \beta \dot{\theta}_0\ ,\quad
I_+^1+iI_+^2=ie^{i(t_0 - \theta_0)}\{\gamma(\dot{t}_0 + \dot{\theta}_0) + i \delta r_1\}\ ,\\
J_-^0&=\alpha \dot{t}_0 - \beta \dot{\theta}_0\ ,\quad
J_-^1+iJ_-^2=e^{-i(t_0 + \theta_0)}\{-\gamma(\dot{t}_0 - \dot{\theta}_0) - i \delta r_1\}\ ,\\
I_-^0&=-\alpha \dot{t}_0 - \beta \dot{\theta}_0\ ,\quad
I_-^1+iI_-^2=ie^{i(t_0 - \theta_0)}\{\gamma(\dot{t}_0 + \dot{\theta}_0) - i \delta r_1\}\ .
 \end{split}
\end{equation}
We cannot find any linear relations between the $''+''$ and $''-''$ sectors. There is no indication of integrability for NDN.

\subsection*{NDD}
For $t_1=\dot{r}_0=\dot{\theta}_0=0$, Eqs.(\ref{JpIpbdry}) and (\ref{JmImbdry}) become
\begin{equation}
 \begin{split}
J_+^0&=\alpha \dot{t}_0 - \beta \theta_1\ ,\quad
J_+^1+iJ_+^2=e^{-it_0}\{-\gamma(\dot{t}_0 -\theta_1) + i \delta r_1\}\ ,\\
I_+^0&=-\alpha \dot{t}_0 - \beta \theta_1\ ,\quad
I_+^1+iI_+^2=ie^{it_0}\{\gamma(\dot{t}_0 +\theta_1) + i \delta r_1\}\ ,\\
J_-^0&=\alpha \dot{t}_0 + \beta \theta_1\ ,\quad
J_-^1+iJ_-^2=e^{-it_0}\{-\gamma(\dot{t}_0 +\theta_1) - i \delta  r_1\}\ ,\\
I_-^0&=-\alpha \dot{t}_0 + \beta\theta_1\ ,\quad
I_-^1+iI_-^2=ie^{it_0}\{\gamma(\dot{t}_0 -\theta_1) - i \delta  r_1\}\ .
 \end{split}
\end{equation}
Again, we cannot find any linear relations between the $''+''$ and $''-''$ sectors. There is no indication of integrability for NDD.

\subsection*{DNN}
For $\dot{t}_0=r_1=\theta_1=0$, Eqs.(\ref{JpIpbdry}) and (\ref{JmImbdry}) become
\begin{equation}
 \begin{split}
J_+^0&=\alpha t_1 - \beta \dot{\theta}_0\ ,\quad
J_+^1+iJ_+^2=e^{-i \theta_0}\{-\gamma(t_1 - \dot{\theta}_0) + i \delta \dot{r}_0\}\ ,\\
I_+^0&=-\alpha t_1 - \beta \dot{\theta}_0\ ,\quad
I_+^1+iI_+^2=ie^{-i\theta_0}\{\gamma(t_1 + \dot{\theta}_0) + i \delta \dot{r}_0 \}\ ,\\
J_-^0&=-\alpha t_1 - \beta \dot{\theta}_0\ ,\quad
J_-^1+iJ_-^2=e^{-i\theta_0}\{-\gamma(-t_1 - \dot{\theta}_0) + i \delta \dot{r}_0\}\ ,\\
I_-^0&=\alpha t_1 - \beta \dot{\theta}_0\ ,\quad
I_-^1+iI_-^2=ie^{-i\theta_0}\{\gamma(-t_1 + \dot{\theta}_0) + i \delta \dot{r}_0 \}\ .
 \end{split}
\end{equation}
We find linear relations between the $''+''$ and $''-''$ sectors as
\begin{equation}
\begin{split}
J_+^A&=R^A{}_B I_-^B\ ,
\quad 
R=
\begin{pmatrix}
1 & 0 & 0 \\
0 & 0 & 1 \\
0 & -1 & 0
\end{pmatrix}
\ ,
\\
I_+^A&=R^A{}_B J_-^B\ ,
\quad 
R=
\begin{pmatrix}
1 & 0 & 0 \\
0 & 0 & -1 \\
0 & 1 & 0
\end{pmatrix}
\ .
\end{split}
\end{equation}
However, for both cases, $R$ are not symmetric. 
There is no indication of integrability for DNN.

\subsection*{DND}
Substituting $\dot{t}_0=r_1=\dot{\theta}_0=0$ into Eqs.(\ref{JpIpbdry}) and (\ref{JmImbdry}), we have
\begin{equation}
 \begin{split}
J_+^0&=\alpha t_1 - \beta \theta_1\ ,\quad
J_+^1+iJ_+^2=\{-\gamma(t_1 -\theta_1) + i \delta \dot{r}_0\}\ ,\\
I_+^0&=-\alpha t_1 - \beta \theta_1\ ,\quad
I_+^1+iI_+^2=i\{\gamma(t_1 +\theta_1) + i \delta \dot{r}_0 \}\ ,\\
J_-^0&=-\alpha t_1 + \beta \theta_1\ ,\quad
J_-^1+iJ_-^2=\{-\gamma(-t_1 +\theta_1) + i \delta \dot{r}_0\}\ ,\\
I_-^0&=\alpha t_1 + \beta \theta_1\ ,\quad
I_-^1+iI_-^2=i\{\gamma(-t_1 -\theta_1) + i \delta \dot{r}_0 \}\ .
 \end{split}
\end{equation}
We find linear relations between the $''+''$ and $''-''$ sectors as
\begin{equation}
\begin{split}
J_+^A&=R^A{}_B J_-^B\ ,
\quad 
R=\textrm{diag}(-1,-1,1)\ ,\\
I_+^A&=R^A{}_B I_-^B\ ,
\quad 
R=\textrm{diag}(-1,1,-1)\ .
\end{split}
\end{equation}
$R$ is symmetric and orthogonal. Also $T'{}^A=R^A{}_B T^B$ gives an automorphism. Therefore, DND is integrable.

\subsection*{DDN}
Substituting $\dot{t}_0=\dot{r}_0=\theta_1=0$ into Eqs.(\ref{JpIpbdry}) and (\ref{JmImbdry}), we have
\begin{equation}
 \begin{split}
J_+^0&=\alpha t_1 - \beta \dot{\theta}_0\ ,\quad
J_+^1+iJ_+^2=e^{-i \theta_0}\{-\gamma(t_1 - \dot{\theta}_0) + i \delta r_1\}\ ,\\
I_+^0&=-\alpha t_1 - \beta \dot{\theta}_0\ ,\quad
I_+^1+iI_+^2=ie^{-i\theta_0}\{\gamma(t_1 + \dot{\theta}_0) + i \delta r_1\}\ ,\\
J_-^0&=-\alpha t_1 - \beta \dot{\theta}_0\ ,\quad
J_-^1+iJ_-^2=e^{-i\theta_0}\{-\gamma(-t_1 - \dot{\theta}_0) - i \delta  r_1\}\ ,\\
I_-^0&=\alpha t_1 - \beta \dot{\theta}_0\ ,\quad
I_-^1+iI_-^2=ie^{-i\theta_0}\{\gamma(-t_1 + \dot{\theta}_0) - i \delta r_1\}\ .
 \end{split}
\end{equation}
We cannot find any linear relations between the $''+''$ and $''-''$ sectors. There is no indication of integrability for DDN.

\subsection*{DDD}
Substituting $\dot{t}_0=\dot{r}_0=\dot{\theta}_0=0$ into Eqs.(\ref{JpIpbdry}) and (\ref{JmImbdry}), we have 
\begin{equation}
 \begin{split}
J_+^0&=\alpha t_1 - \beta \theta_1\ ,\quad
J_+^1+iJ_+^2=\{-\gamma(t_1 -\theta_1) + i \delta r_1\}\ ,\\
I_+^0&=-\alpha t_1 - \beta \theta_1\ ,\quad
I_+^1+iI_+^2=i\{\gamma(t_1 +\theta_1) + i \delta r_1\}\ ,\\
J_-^0&=-\alpha t_1 + \beta \theta_1\ ,\quad
J_-^1+iJ_-^2=\{-\gamma(-t_1 +\theta_1) - i \delta  r_1\}\ ,\\
I_-^0&=\alpha t_1 + \beta \theta_1\ ,\quad
I_-^1+iI_-^2=i\{\gamma(-t_1 -\theta_1) - i \delta r_1\}\ .
 \end{split}
\end{equation}
We find linear relations between $''+''$ and $''-''$ sectors as
\begin{equation}
J_+^A=R^A{}_B J_-^B\ ,\quad I_+^A=R^A{}_B I_-^B\ ,\quad R^A{}_B=-\delta^A_B\ .
\end{equation}
$R$ is symmetric and orthogonal. But $T'{}^A=R^A{}_B T^B$ is not an automorphism. 
There is no indication of integrability for DDD.

\bibliography{bunken_F1_boundary}

\end{document}